\documentclass[aps,pra,twocolumn,showpacs,floatfix,footinbib,groupedaddress,superscriptaddress]{revtex4}
\usepackage{amsmath}
\usepackage{amssymb,graphics,graphicx,times,bm,bbm,multirow}
\usepackage{footnote}
\usepackage{appendix}

\begin{document}

\title{Energy-scales convergence for optimal and robust quantum transport in photosynthetic
complexes}

\author{M. Mohseni}
\affiliation{Center for Excitonics, Research Laboratory of
Electronics, Massachusetts Institute of Technology, Cambridge, MA
02139}
\author{A. Shabani}
\affiliation{Department of Chemistry, Princeton University,
Princeton, New Jersey 08544}
\author{S. Lloyd}
\affiliation{Department of Mechanical Engineering, Massachusetts
Institute of Technology, Cambridge, MA 02139}
\author{H. Rabitz}
\affiliation{Department of Chemistry, Princeton University,
Princeton, New Jersey 08544}

\begin{abstract}
Underlying physical principles for the high efficiency of excitation
energy transfer in light-harvesting complexes are not fully
understood. Notably, the degree of robustness of these systems for
transporting energy is not known considering their realistic
interactions with vibrational and radiative environments within the
surrounding solvent and scaffold proteins. In this work, we employ
an efficient technique to estimate energy transfer efficiency of
such complex excitonic systems. We observe that the dynamics of the
Fenna-Matthews-Olson (FMO) complex leads to optimal and robust
energy transport due to a convergence of energy scales among all
important internal and external parameters. In particular, we show
that the FMO energy transfer efficiency is optimum and stable with
respect to the relevant parameters of environmental interactions and
Frenkel-exciton Hamiltonian including reorganization energy
$\lambda$, bath frequency cutoff $\gamma$, temperature $T$, bath
spatial correlations, initial excitations, dissipation rate,
trapping rate, disorders, and dipole moments orientations. We
identify the ratio of $\lambda T/\gamma\*g$ as a single key
parameter governing quantum transport efficiency, where g is the
average excitonic energy gap.

\end{abstract}

\maketitle

\section{Introduction}

Life on the earth has been solar-powered via the mechanism of
photosynthesis for four billions years \cite{Blankenship02}.
Photosynthetic antenna complexes have evolved to harvest the sun's
energy and efficiently transport it to reaction centers where it is
stored as biochemical energy. The first few steps in photosynthesis
represent highly sophisticated energy capture and transfer
processes, as in an efficient solar cell. Specialized pigments in the
antenna complexes absorb energy from sunlight, creating
electron-hole pairs known as \emph{excitons}. The excitons then
travel to reaction centers where their energy is converted and
stored as chemical energy. The key feature for the success of
photosynthesis is the high efficiency of exciton transport -- as
high as $99\%$ in certain bacterial systems \cite{Chain77}. Thus,
photosynthetic complexes provide an excellent model for designing
efficient artificial excitonic devices. Unfortunately, the
fundamental structural and dynamical processes contributing to such
efficient migration of excitons are not fully understood.

During the last two decades, many spectroscopic observations for
extended excitation states over multiple chromophores within
light-harvesting complexes have been reported
\cite{Grondelle04,Cho05}. These results support the quantum
mechanically delocalized nature of excitations in spatial
coordinate. A variety of theoretical models have been proposed to
capture the nature of delocalized quantum coherence effects in
excitonic energy transfer
\cite{GroverSilbey71,HakenStrobl73,KenkreBook82,
Damjanovi97,Ritz,ZhangMukamel98,Scholes00,Scholes01,YangFleming02,MayBook,Jang04,Renger06}.
These models employ various generalizations of F\"{o}rster energy
transport \cite{Forster65,Scholes03} and Redfield theory
\cite{Redfield65} in different perturbative limits. Specific effects
of quantum coherence in multichromophoric donor to acceptor energy
transfer rates were also studied in details \cite{Jang04}. Recent
advances in multidimensional electronic spectroscopy provide
evidence that long-lived quantum dynamical coherence can exist in
exciton basis
\cite{Engel07,Lee07,Calhoun09,Mercer09,Scholes09-1,Scholes09-2,panit10,Engel12,Engel12-2}.
These experiments reported oscillatory coherences that were observed
to last around $500$ $fs$, which is on the same order as the
transport time-scale. The observations were made at cryogenic and
physiological temperatures for several light-harvesting complexes
including the FMO complex of green sulfur bacteria \cite
{Engel07,panit10}, reaction center (RC) of purple bacteria
\cite{Lee07}, light-harvesting complex II of higher plants
\cite{Calhoun09}, conjugated polymers \cite{Scholes09-1}, and marine
algae \cite{Scholes09-2}. It was specifically suggested by Engel and
Fleming that quantum dynamical interference effects might explain
exceptionally high energy transfer efficiency of light-harvesting
systems \cite{Engel07}. This conjecture has lead to vigorous
theoretical efforts to try to understand the role of coherence in
excitonic transport
\cite{mohseni-fmo,Rebentrost08-1,Rebentrost08-2,Plenio08-1,Plenio09,Castro08,Ishizaki09,Ishizaki09-2,AkiPNAS,CaoSilbey,Hoyer,Caruso10,Sarovar,masoud-tomography,Lloyd10,Abasto12,QEBbook}.
Despite considerable progress in experiment and theory, it is still
unknown how quantum coherence can be preserved in such unusually
warm, complex and wet conditions. Moreover, it is not yet fully
understood if quantum interference effects, in either spatial or
energy coordinates, are contributing to efficiency of these systems.
More importantly, it is not known how robust and optimal are theses
complexes with respect to variations in the coherent system
evolution and/or environmental parameters.

Reference \cite{mohseni-fmo} demonstrated that an effective
collaboration between coherent quantum evolution and environmental
fluctuations could enhance photosynthetic energy transfer efficiency
(ETE). This work was based on a quantum trajectory picture in site
basis \cite{mohseni-fmo}, within the Born-Markov and secular
approximations which guarantee complete positivity of the excitonic
dynamics \cite{BreuerBook}. However, due to the perturbative
approximation used in this model, it was impossible to explore the
optimality of energy transfer efficiency in the relevant regimes of
intermediate system-bath coupling strength. Moreover, due to secular
approximation, the coherence and population transfers are
essentially decoupled. In order to avoid such problems subsequent
studies relied on non-perturbative Haken-Strobl method
\cite{HakenStrobl73} to be able to explore a wider range of
environmental interactions in the context of a (pure-dephasing)
classical white-noise model at infinite temperature. These model
studies illustrate optimal ranges of dephasing-assisted excitation
transfer \cite{Rebentrost08-2,Plenio08-1,Plenio09}. In the optimal
regime, an appropriate level of environmental fluctuations can wash
out the quantum localization effects at the equilibrium state, when
they are not too strong to lead to quantum Zeno effect
\cite{Rebentrost08-2}. These models, however, are by construction
inadequate to capture the role of quantum interplay of system
evolution with non-equilibrium dynamics of bath within realistic
non-perturbative and non-Markovian regimes \cite{Ishizaki09}.

Recently Ishizaki and Fleming developed a general approach for
studying excitation energy transfer in multichromophoric systems
based on a hierarchy of coupled master equations
\cite{Ishizaki09-2}. Using this Hierarchy Equation of Motion (HEOM),
first introduced by Kubo and Tanimura \cite{Tanimura}, quantum
coherent beating in the FMO protein at physiological temperature was
investigated theoretically \cite{AkiPNAS}. Although, HEOM can be in
principle applied to all the different regimes of environmental
interactions as a general benchmark, the requires computational
resources significantly increase with the size of the system, the
bath correlation time-scale, and within low temperature regimes.
Thus, a number of alternative or complementary techniques for
simulation of open quantum dynamics have been recently proposed
\cite{Cao,Jang08,Shi,Nazir09,Plenio10,Liang10,Fujisaki10}. We have
recently shown that the second-order time-convolution (TC2) master
equation can be used to efficiently estimate ETE in large complex
excitonic systems interacting with bosonic environments in the
\emph{intermediate} regimes \cite{Shabani11}. This method is based
on the Cao's earlier work on the generalized Bloch-Redfield master
equations \cite{Cao}. In Ref. \cite{Shabani11} we examined the TC2
master equation reliability beyond extreme Markovian and
perturbative limits.

In this paper, we observe that the FMO pigment-protein complex is
optimal and robust with respect to all the estimated parameters of
system and environmental interactions, within the TC2 master
equation evolution for either of Lorentzian or Ohmic spectral
densities of the environment. The paper contains the following main
results:

\bigskip\noindent{$\bullet$} We show that efficient and fault-tolerant
energy transport occurs when the FMO internal parameters are in tune
with environmental parameters leading to a collaborative interplay
of the coherent free Hamiltonian evolution and incoherent effects
due to environment.

\bigskip\noindent{$\bullet$} We comprehensively study the
effects of reorganization energy $\lambda$, bath frequency cutoff
$\gamma$, temperature $T$, positive and negative spatial
correlations, loss, and trapping mechanisms on the energy transfer
efficiency (ETE) landscapes. We show that ETE is optimal for the
physiologically estimated value of these parameters, and is robust
over a wide range of variations in these parameters.

\bigskip\noindent{$\bullet$}
We demonstrate that a convergence of time/energy scales for the
relevant internal end environmental parameters of FMO complex
facilitates efficient energy transport. For an excitonic system with
average energy gap $g$, we observe that a single effective parameter
$\Lambda = \lambda T/\gamma\*g$ governs the behavior of quantum
transport dynamics. Small values $\Lambda \ll 1$ give rise to weak
localization and low efficiency. Intermediate values $\Lambda\approx
1$ correspond to the optimal energy transfer. Large values $\Lambda
\gg 1$ give rise to strong localization and low efficiency.

\bigskip\noindent{$\bullet$}
We observe that positive/negative spatial correlations essentially
renormalize the reorganization energy to effective lower/higher
values. The positive bath correlations can significantly enhance ETE
at the regime of large $\lambda$ by inducing symmetries in the
effective phonon-exciton Hamiltonian protecting the transport
against strong dynamical disorder. In the intermediate values of
system-bath coupling, the spatial correlations can enhance the
robustness of transport efficiency. However, at the very small
values of reorganization energy, they have an adversarial effect by
diminishing useful but very weak bath fluctuations.

\bigskip\noindent{$\bullet$}
To explore the dynamical role of coherent system evolution, we
examine whether or not the spatial structure of the FMO complex
plays any important role in its performance. We find that the FMO
compact structure is instrumental for supporting high efficiency.
Moreover, we directly demonstrate that the FMO performance is robust
to variations in its geometrical parameters considering its rich
internal parameter space. Specifically, the efficiency of FMO is
robust to variations in the orientations of dipole moments and site
energies due to its compact structure.

\bigskip\noindent{$\bullet$}
Finally, we discuss possibility of a so-called \emph{Goldilocks
principle} in the quantum regime that could explain the convergence
of time-scales in FMO complex and other light-harvesting systems,
with respect to internal parameters and environmental interactions.
The Goldilocks principle for complex systems postulates a `just
right' level of complexity -- too little complexity compromises
function, while too much complexity compromises robustness.
Convergence of timescales increases complexity by allowing different
quantum processes to interact strongly with each other.  These
interactions have the potential either to reduce or to increase
efficiency and robustness.  For the FMO complex and other simplified
small-size light-harvesting systems, we show that the convergence of
timescales effectively acts to increase both efficiency and
robustness. Such a fundamental principle could also serve a key
guiding rule for designing optimal materials to achieve exceptional
quantum transport performance in realistic environments.

\bigskip

The organization of the paper is as follows: section II reviews the
TC2 master equation that we employ to efficiently simulate energy
transport in multichromophoric systems in low excitation regimes
\cite{Shabani11}. In the section III, the efficiency and sensitivity
of the FMO is studied with respect to the environmental parameters
including reorganization energy, bath cutoff frequency, temperature,
bath spatial correlations, and trapping. In section IV, we consider
the dependence of ETE on variations in temporal and geometrical
factors of the trapping mechanism for FMO. The role of initial
excitonic states is investigated in the section V. In the section
VI, we study the robustness of ETE in presence of disorder in the
FMO internal structure parameterized by site energies,
inter-chlorophyl
distances, and dipole moment orientations. 
Some complementary materials for FMO complex are presented in the
appendices on the FMO structural data and ETE in the presence of an
Ohmic bath.

\section{Theoretical model of the FMO complex}

\begin{figure}[tp]
\includegraphics[width=9cm,height=6cm]{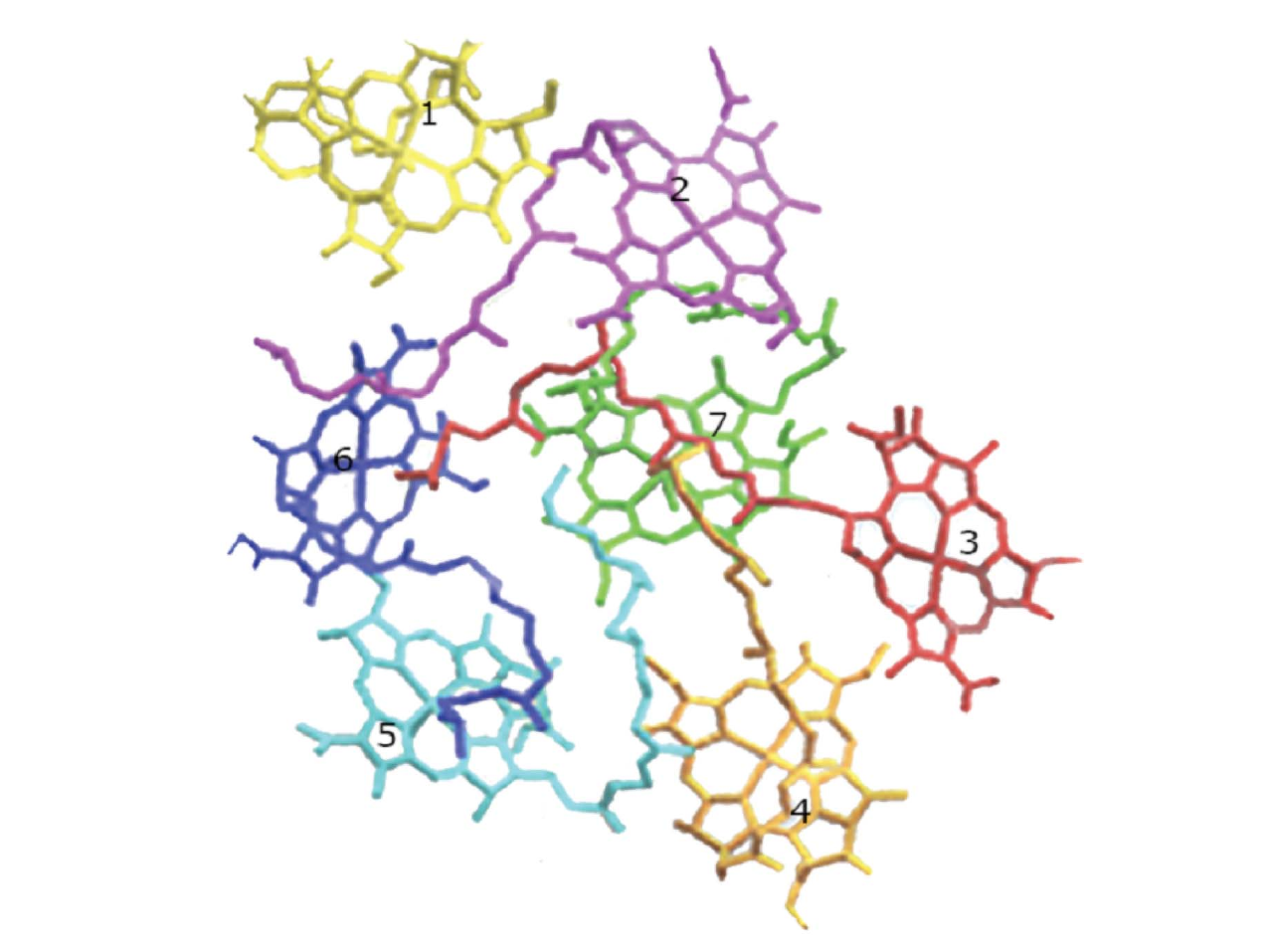}
\caption{The disordered structure of the Fenna-Matthews-Olson (FMO)
complex: It is a trimer consisting of three identical monomers each
formed from seven BChls embedded in a scaffold protein. The FMO
complex acts as an energy transfer channel in green sulphur bacteria
guiding excitons from the light-harvesting antenna complex, in the
proximity of BChls 1 and 6, to the reaction center which is in the
proximity of BChls 3 and 4.} \label{Figfmo}
\end{figure}

To explore optimality and robustness of natural or artificial
light-harvesting complexes we compute the energy transfer efficiency
landscape as a function of the independent degrees of freedom of the
pigment-protein complex including thermal and vibrational
environments. This imposes two fundamental obstacles to developing
analytical and computational approaches that can be employed in this
context. The first problem is the fact that these systems are
complex and contain a significant number of important physical
variables. These independent parameters and their dynamical
interplay have to be fully explored as they could play a major role
in the energy transfer dynamics within these complexes. The second
problem is that the common theoretical models, including
F\"{o}rster, Haken-Strobl, Redfield or Lindblad theory, are based on
simplifying perturbative and/or Markovian assumptions that are
inadequate by construction \cite{Ishizaki09}. This second problem
exists even if the estimated parameters of the system or system-bath
couplings are rather weak and when bath has little or no memory. The
difficulty is that for studying the optimality or robustness of
these complexes one has to fully explore both topology and geometry
of energy transfer efficiency landscape outside of the known values
of a given pigment-protein complex \cite{landscape}. The combination
of these two problems indicates that such study will be practically
intractable with current state of the art simulation techniques such
as HEOM \cite{AkiPNAS}, which, while accurate, are computationally
highly expensive. Recently, we have derived the well-known time
nonlocal master equation TC2 without making the usual weak
system-bath coupling assumption. In this section, we summarize the
main relevant definitions and equations of that approach. For more
technical details we refer the readers to Ref. \cite{Shabani11}.

The dynamics of a multichromophoric system interacting with
surrounding scaffold protein and solvent can be understood by
starting from a general time evolution formulation of open quantum
systems. The total system-bath Hamiltonian can be expressed as
\begin{equation}
H_{total}=H_{S}+H_{ph}+H_{S-ph}\label{Hamiltonian}
\end{equation}%
where
\begin{eqnarray*}
H_{S} &=&\sum_{j,k}\epsilon _{j}|j\rangle \langle j|+J_{jk}|j\rangle
\langle k|, \\
H_{ph} &=&\sum_{j,\xi }\hbar \omega _{\xi }(p_{j,\xi }^{2}+q_{j,\xi
}^{2})/2, \\
H_{S-ph} &=&\sum_{j}S_{j}B_{j}.
\end{eqnarray*}%
The phonon bath is modeled as a set of harmonic oscillators
\cite{McKenzie}. Here $|j\rangle $ denotes an excitation state in a
chromophore spatially located at site $j$. The diagonal site
energies are denoted by $\epsilon _{j}$s that include
 reorganization energy shifts $\lambda _{j}=\sum_{\xi }\hbar
\omega _{\xi }d_{j,\xi }^{2}/2$ due to interactions with a phonon
bath; $d_{j,\xi}$ is the dimensionless displacement of the
$(j,\xi)$th phonon mode from its equilibrium configuration. The
strengths of dipole-dipole interactions between chromophores in
different sites are represented by $J_{jk}$. The operators
$S_{i}=|j\rangle \langle j|$ and $B_{j}=-\sum_{\xi }\hbar \omega
_{\xi }d_{j,\xi }q_{j,\xi }$ are system and bath operators. Here, we
assume that each site is linearly interacting with a separate phonon
bath. The overall dynamics of the system can be expressed by a
time-nonlocal master equation, e.g., TC2, \cite{Shabani11} as:
\begin{eqnarray}
&&\frac{\partial}{\partial t}\rho(t)=\mathcal{L}_S\rho(t)+\mathcal{L}_{e-h}\rho(t) \label{TNME}\\
&&-\sum_j[S_{j},\frac{1}{\hbar^2}\int_0^t C_j(t-t')
e^{\mathcal{L}_S(t-t')} S_{j}\rho(t') dt'-h.c.]\notag
\end{eqnarray}
where $C_{j}(t-t_{1})=\langle \tilde{B}_j(t)\tilde{B}_j%
(t_{1})\rangle $ represent the bath correlation function, and the Liouvillian superoperators $\mathcal{L}%
_{S}$, $\mathcal{L}_{ph}$ and $\mathcal{L}_{S-ph}$ associated to
$H_{S}$, $H_{ph}$ and $H_{S-ph}$ respectively. The term
$\mathcal{L}_{e-h}=-\sum_j r_{loss}^j\{|j\rangle\langle j|,.\}-
r_{trap}\{|trap\rangle\langle trap|,.\}$ captures two different
competing electron-hole pair recombination processes that determine
the energy transfer efficiency of light harvesting complexes. The
first process is the loss due to dissipation to the environment at
each site that happens within the time-scale of $1$ ns. This
inevitable adverse environmental effect guarantees that the energy
transfer efficiency has a value less than one. The second process is
the desired recombination process due to successful trapping.

The ETE is defined as accumulated probability of exciton being
successfully trapped:

\begin{eqnarray}
\eta=2r_{trap}\int_0^\infty \langle trap|\rho(t) |trap\rangle dt
\label{ETE}
\end{eqnarray}

The above performance function is biologically relevant and has been
extensively used for a variety of light-harvesting complexes
\cite{Ritz,mohseni-fmo,Shabani11}.

\begin{figure}[tp]
\includegraphics[width=9cm,height=6cm]{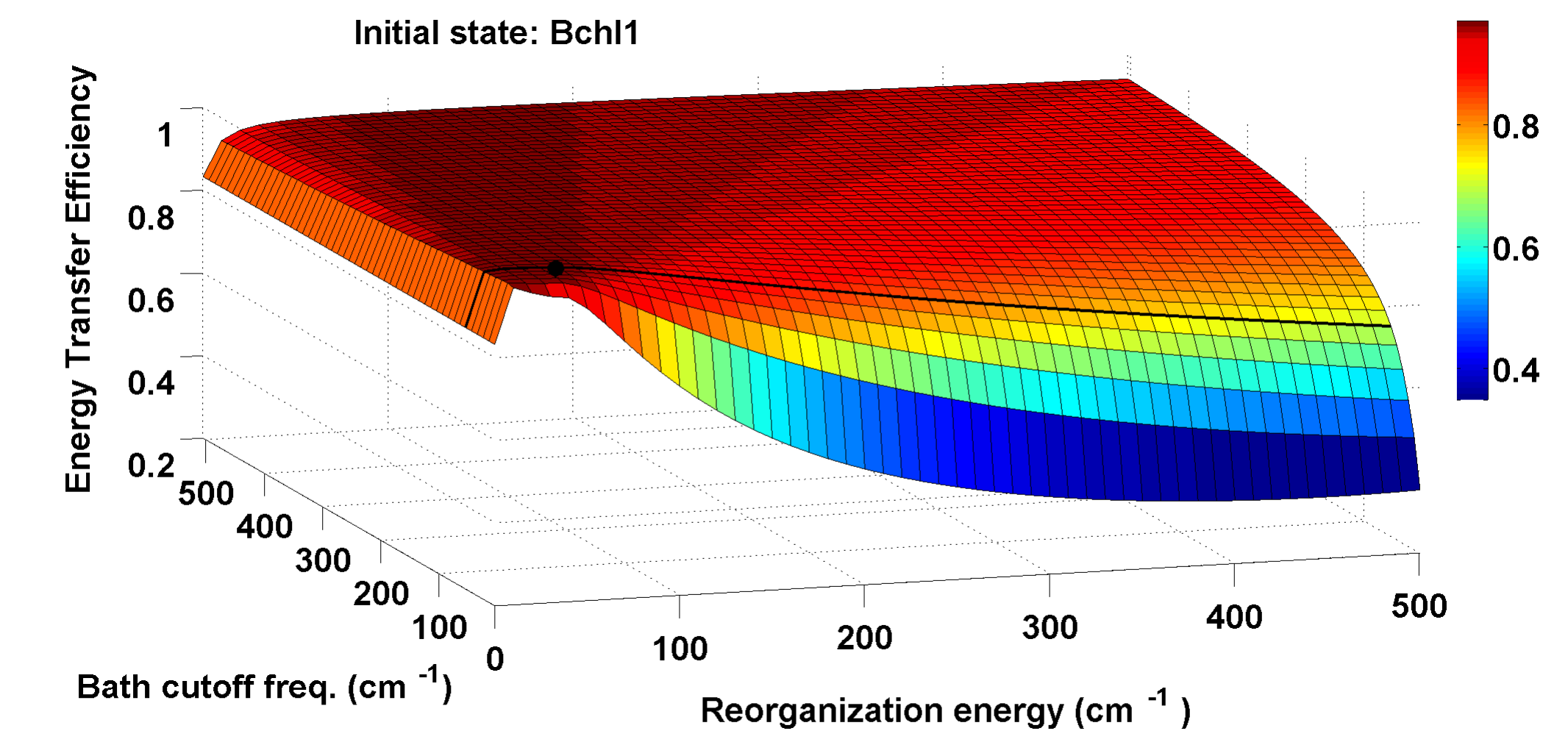}
\caption{The Energy transfer efficiency (ETE) of the
Fenna-Matthews-Olson complex versus reorganization energy $\lambda$
(as a measure of decoherence strength) and bath cutoff frequency
$\gamma$ (as a measure of non-Markovian character of the bath
\cite{Patrick,Breuer09}). The experimentally estimated values of
$T=298$ $^\circ K$, $\lambda=35$ $cm^{-1}$, $\gamma=50-166$
$cm^{-1}$, $r_{trap}^{-1}=1$ $ps$ and $r_{loss}^{-1}=1$ $ns$ reside
at an optimal and robust neighborhood of ETE. The FMO complex would
act sub-optimally in the regimes of large $\gamma$ and very small
$\lambda$, or large $\lambda$ and very small $\gamma$. In the
intermediate values of $\lambda/\gamma$ the phenomenon of
environment-assisted quantum transport takes place. A top view of
this plot in Fig. \ref{topview} indicates $\lambda/\gamma$ is the
parameter that governs efficiency at a fixed temperature.}
\label{Figeffgamlam}
\end{figure}

In this work, we mostly concentrate on the dynamics of the FMO
pigment-protein complex of the bacterium Chlorobium tepidum
\cite{FMOxray2} as a prototype for larger light-harvesting antenna
complexes, Fig. \ref{Figfmo}. The structure of this pigment-protein
complex was the first of light-harvesting complexes to be determined
by x-ray crystallography \cite{FMOxray}. The FMO structure consists
of a trimer, formed by three identical monomers, each includes a
closely packed assembly of seven BChl-a molecules. The FMO trimer
guides excitation energy transfer between the chlorosome, the
light-harvesting antennae of green sulfur bacteria, and the
membrane-embedded type I reaction center (RC). In the following
sections for our numerical simulations, unless specified otherwise,
the environmental parameters for the FMO complex are chosen
according to the experimentally or theoretically estimated values of
reorganization energy $35$ $cm^{-1}$, temperature $298$ $^\circ K$,
trapping rate of $1$ $ps$, exciton life-time of $1$ $ns$, and zero
spatial correlations. In the present simulations, we adopt the bath
cutoff frequency of $50$ $cm^{-1}$ from the studies in Refs.
\cite{Fleming96,Cho05}. However we have repeated our simulations for
another reported estimation of $\gamma=166$ $cm^{-1}$, Ref.
\cite{Read08}, and we observed no significant difference in the
behavior of ETE landscape. The diagonal and off-diagonal free
Hamiltonian parameters are given in appendix A, as functions of
chromophoric distances, dipole moment angles and site energies. The
trapping rate is treated as a free parameter in section IV. For most
of this manuscript we assume an excitation initially localized at
site 1. Similar results are obtained for other initializations such
as localized state at site 6. We study the overall, best, and worst
case scenarios of dependency on the initial states in the section V.
In the next section, we use Eq. (\ref{ETE}) to demonstrate
efficiency and robustness of the FMO complex with respect to
variations in reorganization energy, temporal correlations,
temperature and spatial correlations.

\section{optimality and robustness with respect to environmental
parameters}

We employ the time-nonlocal master equation presented above to
efficiently estimate the energy transfer efficiency \emph{landscape}
as a function of various independent system and environmental
degrees of freedom over a wide range of values. This efficient
simulation allows us to examine comprehensively all relevant regimes
of the multiparameter space for finding possible high efficient and
robust neighborhoods. Only after such exhaustive study can one
quantify the performance of any particular natural photosynthetic
complexes. Moreover, such studies shed light on the maximum
capabilities that can be achieved for optimal material design to
engineer and characterize fault-tolerant artificial light-harvesting
systems \cite{Francis08,FrancisJACS10} within a given rich
system-environmental parameter space. The quantum efficiency of
photosynthetic energy conversion can also be estimated by measuring
the quantum requirements of ATP formation \cite{Chain77}. Thus, the
overall robustness and optimality of a pigment-protein complex can
also be experimentally explored, verified, or calibrated by varying
the tunable parameters in the laboratory.  For example, this can be
achieved by changing ambient temperature and using diverse solvents
with different dielectric properties \cite{FrancisJACS10}.

\begin{figure}[tp]
\includegraphics[width=9cm,height=6cm]{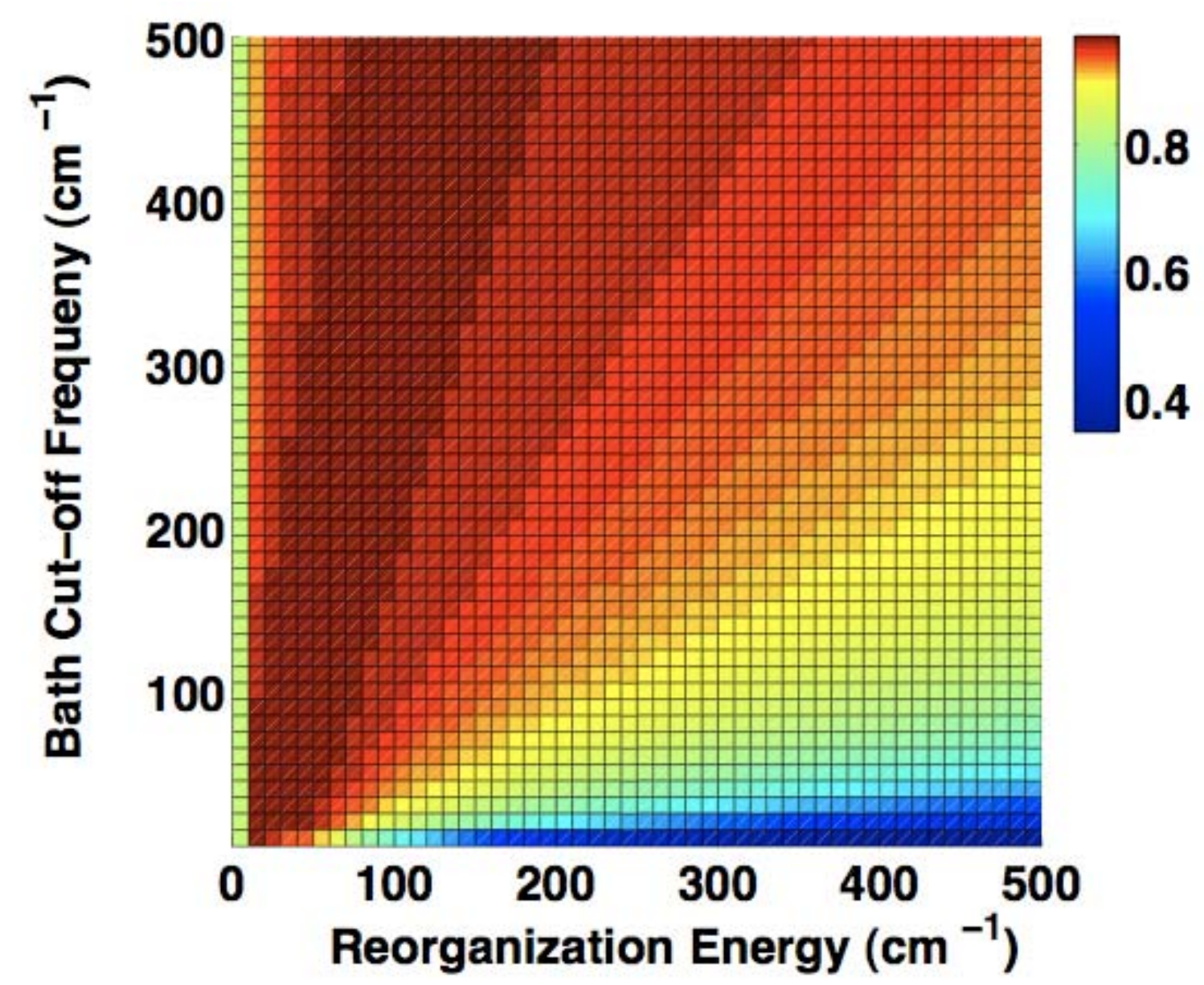}
\caption{Top view of ETE landscape illustrated in Fig.
\ref{Figeffgamlam} indicating that the ratio of the reorganization
energy over bath frequency cutoff can be considered as the parameter
that governs the energy transfer efficiency at a fixed temperature.
As we inspect this plot in an angular coordinate from the vertical
axis, $\gamma$, toward the horizontal axis, $\lambda$, we can
distinguish different regions of the ETE landscape that are
separated by straight lines $\lambda/\gamma$. At very small
decoherence rate, the FMO complex experiences weak localization due
to static disorder. As we increase this ratio, an optimal region of
ETE emerges that is induced by an appropriate level of interplay
between environmental fluctuations and coherent evolution. At higher
levels of this parameter, ETE drops significantly due to strong
localization induced by dynamical disorder.} \label{topview}
\end{figure}

In addition to our landscape study of ETE, we also examine the
degree of optimality and robustness of the energy transfer by
employing the the Euclidean norm of the gradient and Hessian matrix
of the ETE function. The Euclidean norm of the ETE gradient at any
parameters values $p_1$ and $p_2$, $||\nabla{\eta(p_1,p_2)}||_2$,
quantifies the degree of optimality. The gradient measure reveals
the degree of local optimality in a surface manifold. Careful
inspections of the room temperature plots for the ETE function
versus various pairs of relevant parameters show a convex or concave
manifold, thus gradient as a measure of local optimality suffices to
measure global optimality. To examine the robustness, we compute the
Hessian matrix norm $||H(\eta(p1,p2))||
_2=||[\partial^2\eta/\partial p_i\partial p_j]||_2$ ($i,j=1,2$) as
the total  measure of local curvature of the manifold. A smaller
value of this norm corresponds to a flatter surface, thus a more
robust process. We use a five-point stencil method to compute
derivatives numerically.

\subsection{Reorganization energy and non-Markovianity}

We first explore the variation of the FMO energy transfer efficiency
versus reorganization energy and bath cutoff frequency using a
Lorentzian spectral density, see Fig. \ref{Figeffgamlam}. The
reorganization energy, $\lambda$, is proportional to the squared
value of the system-bath couplings, quantifying the decoherence
strength. The bath cutoff frequency is the inverse of the bath
coherence time-scale that captures the non-Markovian nature of the
environment. That is, the non-Markovianity measure, defined as the
information flow from the system to the phonon bath by Breuer et.
al. \cite{Breuer09}, increases exponentially with decreasing bath
cutoff frequency \cite{Patrick}. 
The ETE function has not been illustrated in Fig. \ref{Figeffgamlam}
for bath cutoff frequency values less than $\gamma=5$ $cm^{-1}$ ,
since according to our analysis in Ref. \cite{Shabani11} the
simulation errors of TC2 master equation may become significant in
such highly non-Markovian regimes. It should be noted that in the
regime of $\gamma>\beta^{-1}$ we apply low temperature corrections
to the bath correlation function as explained in the next section.

The optimality and robustness of ETE for the FMO protein complex at
the experimentally estimated values of $\lambda=35$ $cm^{-1}$ and
$\gamma=50-166$ $cm^{-1}$ are evident in Fig. \ref{Figeffgamlam}. An
independent study on the optimality of ETE versus reorganization
energy has been recently reported in Ref. \cite{Jianlan10}. It can
be observed in Fig. \ref{Figeffgamlam} that the non-Markovianity of
the bath can slightly increase ETE in the regimes of weak
system-bath coupling. However, such slow bath behavior can
significantly decrease ETE when the system interacts strongly with
it. The main question is how one can understand this phenomenon for
all non-Markovian and Markovian regimes in the context of
Environment-Assisted Quantum Transport (ENAQT), which was first
investigated using simplified formalisms of the Lindblad (weak
coupling and Markovian assumptions) \cite{mohseni-fmo} and
Haken-Strobl (infinite temperature, pure-dephasing assumption)
\cite{Rebentrost08-1}.

In this work, we report an underlying theory for
environment-assisted quantum transport. The landscape in Fig.
\ref{Figeffgamlam} shows a remarkable interplay of reorganization
energies $\lambda$ and bath frequency cutoff $\gamma$. The ETE takes
values below unity if the FMO operates at the limit of very small
$\lambda$ and large $\gamma$. On the other limit, the FMO efficiency
drops significantly when operating at large $\lambda$ and very small
$\gamma$. As we argue below these two regimes can be understood as
manifestations of weak- and strong quantum localization,
respectively.

A note on terminology: we use the terms weak and strong localization
in analogue to the corresponding effects in bulk solid-state systems
\cite{kramer93}.  However, in contrast to solid state systems, here
we are dealing with finite-size systems of few chromophores. In
addition, it is important to note that because of the long-range
nature of the dipolar force, the bulk versions of these chromophoric
systems would not exhibit exponentially localized states. What is
really being investigated here, then is a kind of {\it transient}
localization which would eventually go away due to the long-range
nature of the dipolar force and the small system size.  If the
destruction of transient localization takes longer than the exciton
lifetime, however, then transient localization is just as
adversarial to quantum transport efficiency
as full localization. 
In this paper, when we refer to weak and strong localization, the
reader should keep in mind that we are actually discussing weak and
strong transient localization.

In order to fully explore the regimes of weak and strong
localization and the various intermediate transitional regions, we
illustrate a top view of Fig. \ref{Figeffgamlam} in Fig.
\ref{topview}.  This plot immediately reveals that the regions of
distinct energy transfer efficiencies at room temperature are
governed by a parameter of the form $\lambda/\gamma$.   For small
and for large values of this parameter, the efficiency is low.  The
efficiency reaches its maximum for intermediate values of this
parameter. In both strong and weak quantum localization limits the
excitation will be spatially trapped in the regions typically far
from reaction center and eventually dissipating to bath due to
adversarial electron-hole recombination processes which occur on
three order of magnitude slower time-scale. In the intermediate
regime, the right amount of the interplay of quantum coherence and
environmental fluctuations can facilitate an optimal energy
transport in a robust fashion by various physical mechanisms
including minimizing site energy mismatches, washing out potential
destructive quantum interference effects, and enhancing the energy
funneling by providing an appropriate vibrational energy sink.

The dependence of efficiency on  $\lambda/\gamma$ can be understood
as follows.  When the FMO complex interacts weakly with its
environment the exciton migration is essentially dictated by the
site energies and inter-pigment couplings. The spatial locations,
Fig. \ref{Figfmo}, and dipole moments of BChls do not show any
obvious regular pattern.   In the absence of environmental
interactions, destructive interference effects due to such random
configurations can cause weak localization over a few chromophores
away from the trapping site. This phenomenon prevents successful
delivery of the initial exciton to the reaction center and therefore
leads to below unity efficiency of about $83\%$ at $\lambda=0$. Weak
localization is amplified by lowering the temperature, a behavior
observed in the simulations presented in the next section. By
increasing the system-bath coupling strength, the adversarial
interference effects of excitonic pathways are diminished in a
fashion similar to observations reported in Refs.
\cite{mohseni-fmo,Rebentrost08-1,Plenio08-1}. However, here it can
be observed in Fig. \ref{topview} that for larger bath cutoff
frequencies, a larger reorganization energy is required for ENAQT to
occur.

We can understand these observations in terms of ENAQT by noting
that the effective decoherence rate is given by $\lambda/\gamma$ in
the perturbative limit at a fixed temperature \cite{BreuerBook}.
This ratio becomes smaller as we raise the $\gamma$ therefore a
stronger coupling $\lambda$ is needed to guarantee the level of
decoherence strength required for ENAQT. The overlap of a
delocalized exciton wave function with the trap enables an almost
complete, $98\%$ quantum transport in the optimal range of $\lambda
/\gamma$ at the ambient temperature. As we increase the
reorganization energy and bath coherence time-scale, the ETE starts
to drop. In this regime, excitonic wave function again experiences
localization as the environmental fluctuations change their role to
play a strong adversarial effect on the quantum transport
essentially as a source of strong dynamical disorder. The ETE
landscape in Fig. \ref{topview}, clearly has level sets that exhibit
linear relationship with $\lambda$ and $\gamma$. Indeed the ratio
$\lambda/\gamma$, known as the Kubo number, is the parameter that
governs Anderson localization transition in stochastic classical
modeling of environmental interactions.
\cite{kramer93,Reineker88,Cast00,Goychuk05}. In the fixed
high-temperature limit of ETE illustrated in Fig. \ref{topview}, one
can observe that $\lambda/\gamma$ is a determining parameter for
transport efficiency in the regions beyond optimal ENAQT area. In
the next section, we go beyond the Kubo number, by directly
investigating the temperature dependent energy transfer dynamics,
leading to a general governing parameter as $\lambda T/\gamma$.

\begin{figure}[tp]
\includegraphics[width=9cm,height=6cm]{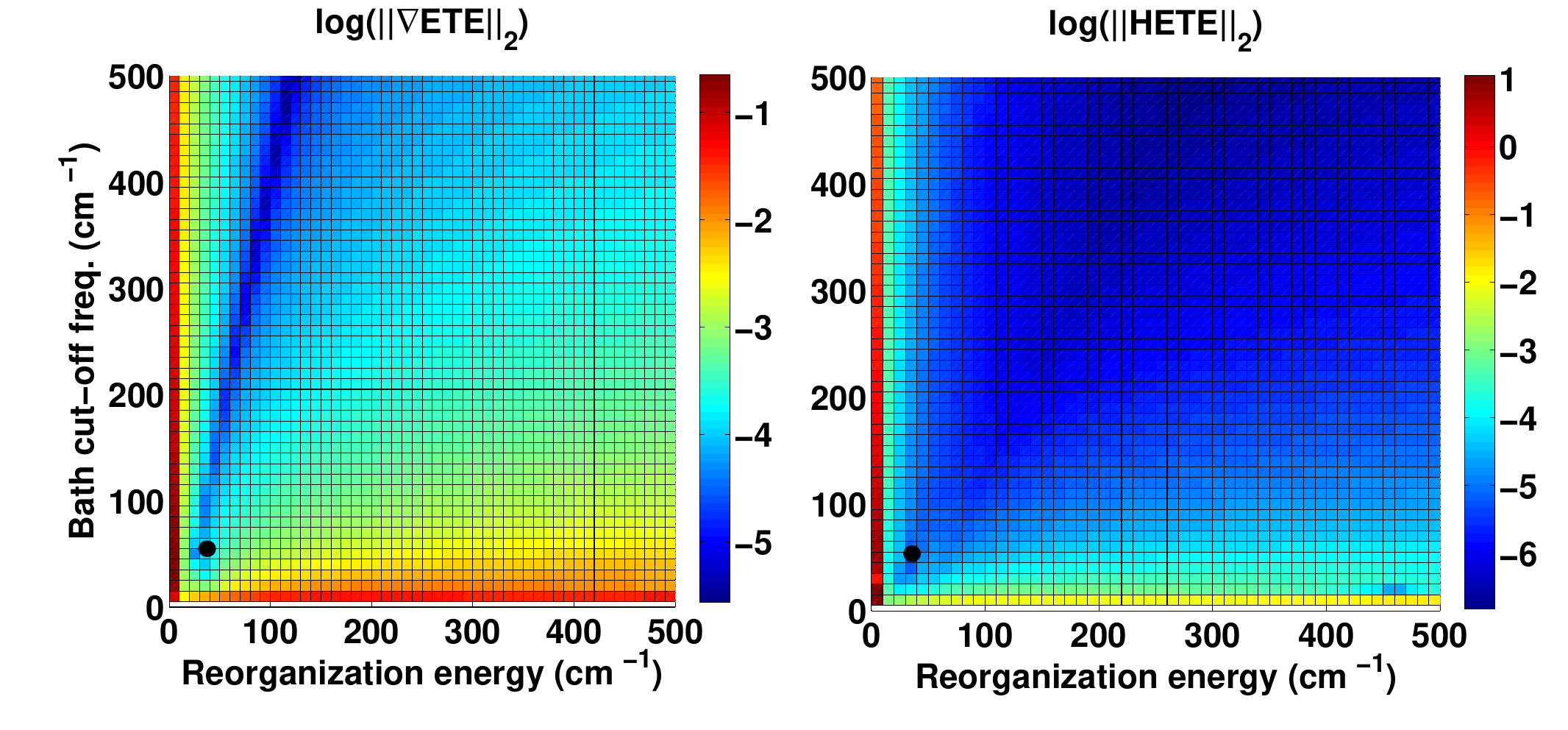}
\caption{This figure is a complement to Fig. \ref{Figeffgamlam}.
(Left) The degree of ETE optimality is quantified at different
values of $\lambda$ and $\gamma$ by the gradient matrix norm of the
ETE function. The dark blue points represent near optimal values.
(Right) The degree of ETE robustness is quantified by the Euclidian
norm of Hessian of the ETE. The dark blue points represent near
robust points. The estimated FMO environmental values of $\lambda=35
cm^{-1}$ and $\gamma=50$ $cm^{-1}$, marked by black dots, are
located on the corner of both robust and optimal region. This
implies the FMO environmental conditions provide the minimal work
required for facilitating optimal and robust energy transport by
maintaining a rather weakly-coupled small scaffold protein.}
\label{Figopt-sus-lamgam}
\end{figure}

A more quantitative study of the degree of optimality and robustness
of the energy transfer as functions of system-bath coupling strength
and bath memory is illustrated in Fig \ref{Figopt-sus-lamgam}. The
optimality is defined as Euclidean norm of the ETE function gradient
$||\nabla{\eta(\lambda,\gamma)}||_2$ to locate the local maxima in
the ETE landscape in Fig. \ref{Figopt-sus-lamgam}. The robustness is
defined by $||H(\eta(\lambda,\gamma))|| _2$ to measure local
curvature of the manifold. Note that the ETE gradient and Hessian
matrix norms are indicated in a logarithmic scale, thus the global
optimal point with zero derivative can not be explicitly highlighted
in this representation. The experimentally estimated values for FMO
are illustrated as black dots in each graph clearly located in an
optimal and robust region. One remarkable feature is the fact that
environmental parameters of FMO have almost the minimal
reorganization energy and bath cutoff frequency among all the
regions with simultaneous optimal and robust properties. This might
be explained by the overall tendency in nature to minimize the
amount of required work, that is, facilitating an optimal and robust
environmental platform for the FMO energy transport by preserving a
rather small size scaffold protein that is weakly coupled.

The theoretical and experimental modeling of the environment
surrounding the FMO complex suggests that the spectral density of
the phonon bath modes can be expressed as a some of a few Lorentzian
terms \cite{McKenzie}. The spectral density function $J(\omega)$
determines the time correlation functions as:
\begin{eqnarray}
\langle\tilde{B}(t)\tilde{B}(0)\rangle=\frac{1}{\pi}\int_0^\infty
d\omega J(\omega)\frac{\exp(-i\omega t)}{1-\exp(-\beta\hbar\omega)},
\end{eqnarray}
where a Lorentzian spectral function has the form
$J(\omega)=2\lambda\omega/(\omega^2+\gamma^2)$. For simplicity in
this work we have considered a single Lorentzian term with amplitude
$\lambda=35$ $cm^{-1}$ and cutoff frequency $\gamma=50$ $cm^{-1}$.
However these values were actually extracted by fitting the
experimentally measured absorption spectrum using an Ohmic spectral
density $J(\omega)=\lambda(\omega/\gamma)\exp(-\omega/\gamma)$
\cite{Fleming96}. Thus, we also examine ETE versus variations of
$\lambda$ and $\gamma$ in such a model depicted in Fig.
\ref{plotohm}, see appendix B. We note that for an Ohmic model a
very similar behavior of the optimality and robustness of energy
transport can be observed as those in Fig. \ref{Figeffgamlam};
however, the estimated ETE is lower than those obtained by
exploiting a Lorentzian model. Specifically for FMO the ETE value
are $92.3\%$ and $96.7\%$ for Ohmic and Lorentzian spectral density
respectively. This suggests that an Ohmic bath can not capture the
near unity efficiency of FMO complex, and favors the Lorentzian
model as a more accurate description of the bath spectral density.

\subsection{Reorganization energy and Temperature}

\begin{figure}[tp]
\includegraphics[width=9cm,height=6cm]{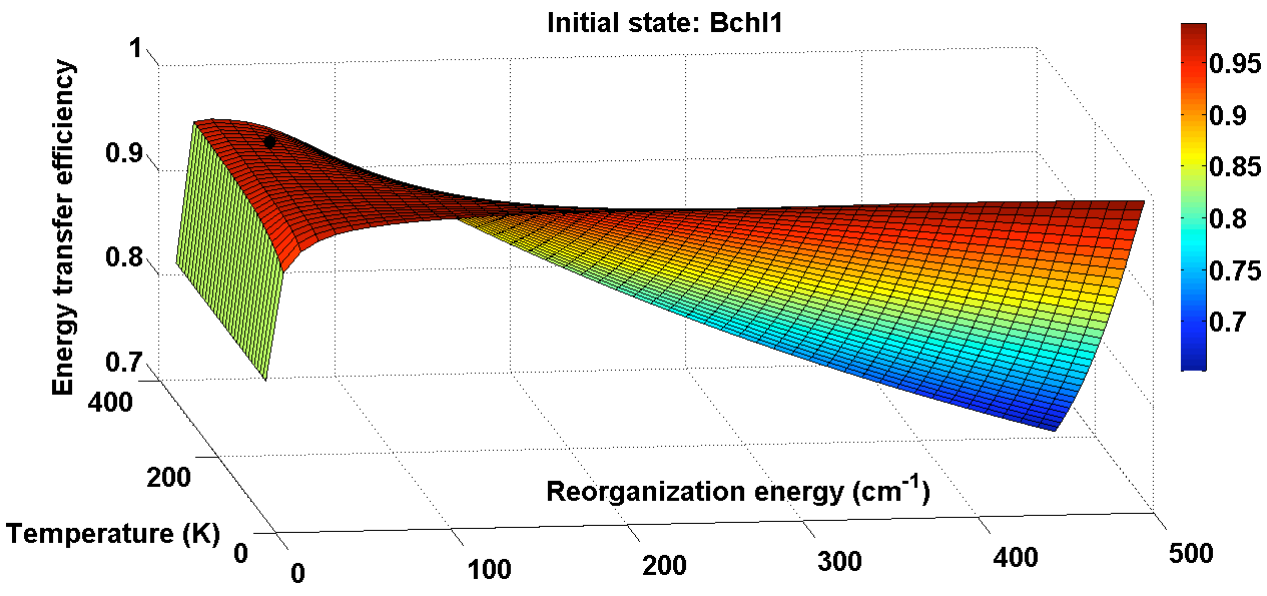}
\caption{Energy transfer efficiency manifold as a function of
reorganization energy and temperature. The temperature range is from
$35$ $^\circ K$ to $350^\circ K$. At the regimes of large
system-bath coupling strength the ETE drops faster by increasing the
temperature due to strong localization induced by dynamical
disorder. There is a narrow region of reorganization energies
between $10$ and $50$ $cm^{-1}$ that ETE is relatively robust with
respect to variations in temperature. The FMO value of $\lambda=35$
$cm^{-1}$ is located in this region indicating a significant
insensitivity to temperature variations. At very low reorganization
energy and low temperatures weak localization can be observed due to
static disorder.} \label{Figefflamtemp}
\end{figure}

Here, we directly examine the temperature dependence of the
effective parameter $\lambda/\gamma$ discussed in the pervious
section. Generally light-harvesting complexes in bacterial, marine
algae, and higher plants operate at various temperatures and
reorganization energies. For example, the FMO complex of the green
sulfur bacteria can operate at the bottom of the ocean at a depth of
hundreds of meters, and also in hot springs with diverse temperature
variations. Moreover, 2D electronic spectroscopy of these systems
have been performed at different cryogenic and room temperatures.
Thus, we need to explore the efficiency and sensitivity of energy
transport as a function of reorganization energy and temperature.

We note that in the low temperature limit, $\beta>\gamma^{-1}$, the
Lorentzian bath correlation function $\langle\tilde{B}(t)\tilde{B}
(t')\rangle _{ph}$ is no longer a single term
$\lambda(2/\beta-i\gamma)e^{-\gamma(t-t')}$ and should be corrected
by a sum of exponentially decaying terms

\begin{eqnarray}
\langle\tilde{B}(t)\tilde{B}(t')\rangle
_{ph}=\frac{\gamma\lambda}{\hbar}(\cot(\frac{\beta\hbar\gamma}{2})-i)e^{-\gamma (t-t')}+\notag\\
\frac{4\lambda\gamma}{\beta\hbar^2}\sum_{k=1}^{\infty}\frac{\nu_k}{\nu_k^2-\gamma^2}
e^{-\nu_k(t-t')}\label{lowT}
\end{eqnarray}
where $\nu_k=2\pi k/\beta\hbar$ are bosonic Matsubara frequencies.
In practice a truncation of the above infinite series is needed for
numerical simulations.  The higher levels of truncation are dictated
by lower temperature limits for the systems under study. To
guarantee an accurate estimation of the bath correlation function
for the calculation of energy transfer landscape, in the relevant
low temperatures, we consider the first one hundred Matsubara
frequencies in the above summation. It should be noted that the
mentioned low temperature corrections have a minor numerical cost
for calculating ETE by using TC2 (\ref{TNME}), since we just need to
compute sum of the Laplace transforms of the above exponential terms
\cite{Shabani11}. This is another advantage of using the TC2 in
contrast to the more accurate approaches such as HEOM for which
every correction term $\nu_k e^{-\nu_k(t-t')}$ necessitates
considering $N$ extra auxiliary variables in the simulation, where
$N$ is the number of the sites, therefore significantly increasing
the computational cost of the simulations.

\begin{figure}[tp]
\includegraphics[width=9cm,height=6cm]{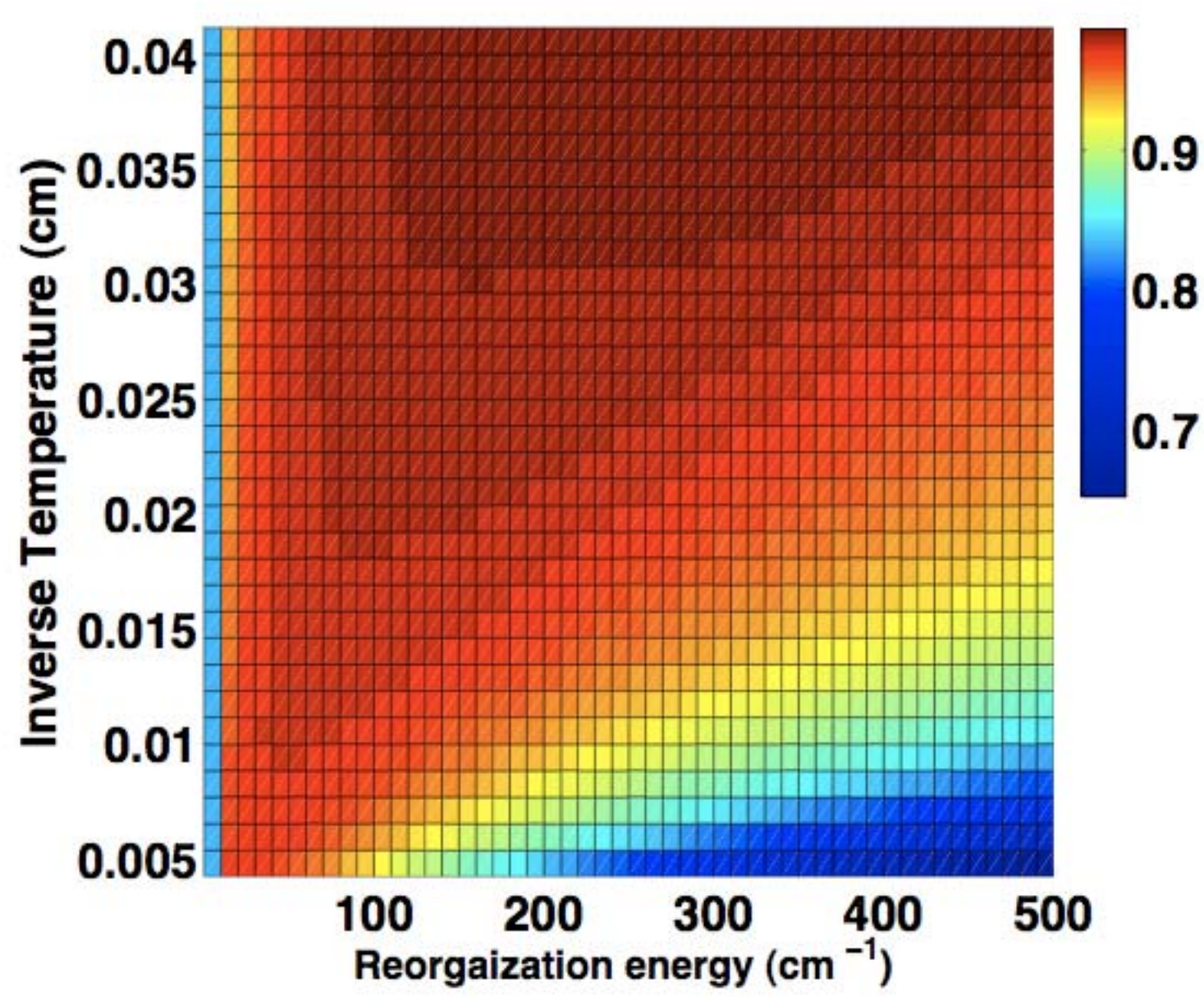}
\caption{The top view of the ETE landscape presented in Fig.
\ref{Figefflamtemp} as a function of inverse temperature and
reorganization energy. In the high temperature regime
($\beta^{-1}>\gamma$), the three regions of weak localization, ENAQT
and strong localization can be distinguished by the parameter
$\lambda T$ given the free Hamiltonian of FMO and $\gamma=50$
$cm^{-1}$. These results, combined with Fig. \ref{topview} suggest
that the parameter $\lambda T/\gamma$ governs the shape of the
overall FMO energy transfer landscape.} \label{topT}
\end{figure}

\begin{figure}[tp]
\includegraphics[width=9cm,height=6cm]{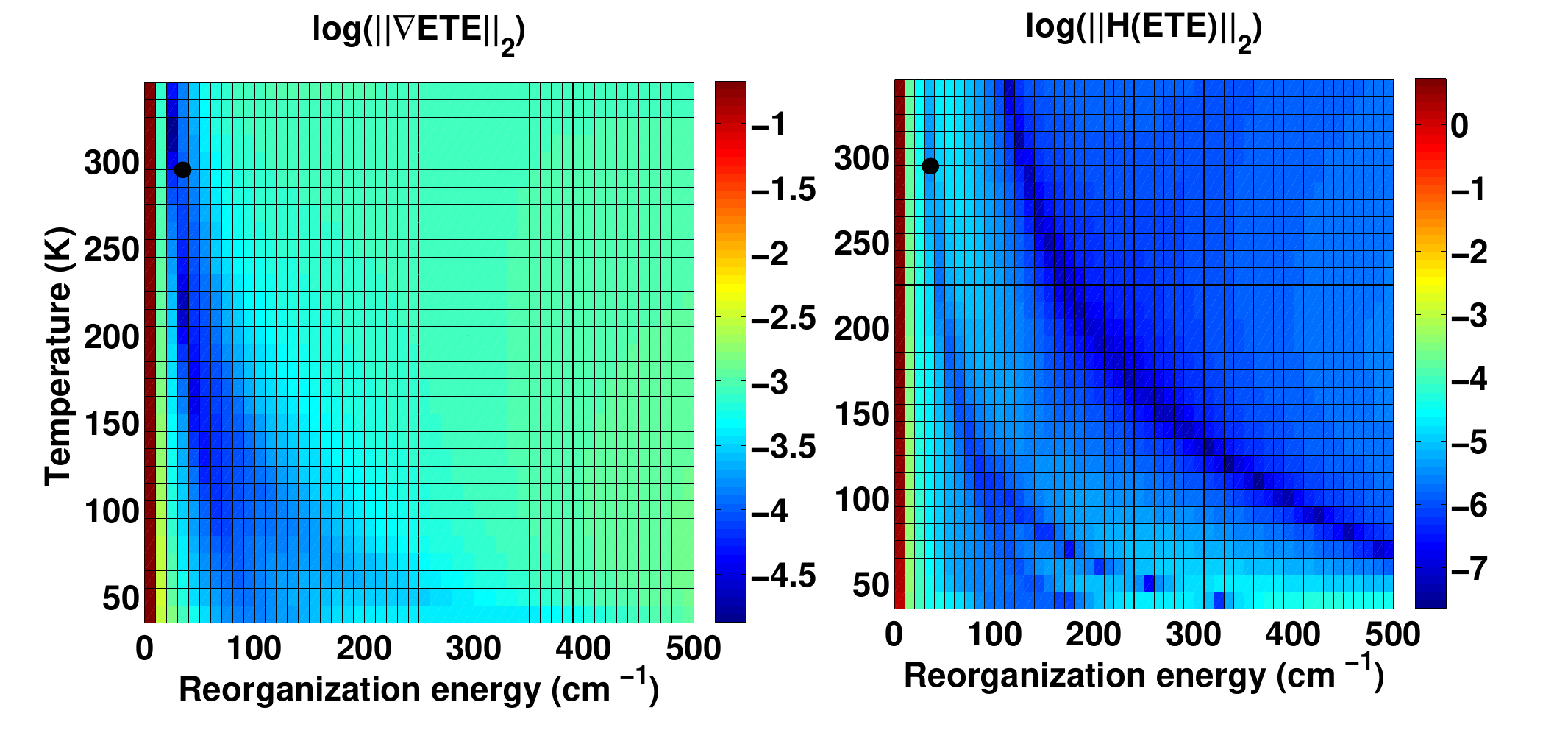}
\caption{This figure is a complement to Fig. \ref{Figefflamtemp}.
Left: the degree of ETE optimality is quantified at different values
of $\lambda$ and $T$ by the norm of the ETE gradient. The dark blue
points represent near optimal values. Right: the degree of ETE
robustness is quantified by the Euclidian norm of the ETE Hessian.
The dark blue points represent near robust points. We note that
within the range of possible FMO operating temperatures (e.,g
between $T=280^\circ K$ to $T=350 ^\circ K$) simultaneous optimal
and robust energy transport can only be achieved for $\lambda$
values around $30$ to $35$ $cm^{-1}$, that is equivalent to the
estimated reorganization energy for the FMO complex that is marked
by a black dot at room temperature $T=298^\circ K$.}
\label{Figopt-sus-lamT}
\end{figure}

We show the behavior of ETE as a function of reorganization energy
$\lambda$ and temperature $T$ in Fig. \ref{Figefflamtemp}. It can be
observed that at various possible conditions for the FMO protein of
$T=280$ to $350$ $^\circ K$ and $\lambda=35$ $cm^{-1}$, the ETE has
an optimal value and resides in a robust region of the energy
transfer landscape. We note that for system-bath strength
corresponding to the FMO's environment, the energy transport is very
robust to variations of temperature and just slightly decreases in
the low temperature limit. By increasing the reorganization energy,
the temperature dependence of ETE becomes more pronounced. As we
move toward the classical regime of the dynamics at high
reorganization energy and high temperature ETE drops significantly.
This implies the necessity of quantum effects for highly efficient
and robust excitonic energy transfer. It is remarkable that at the
relevant physiological temperatures, the ETE is robust only within
the range of $\lambda=35$ $cm^{-1}$, and it becomes unstable as we
approach reorganization energy of over $100$ $cm^{-1}$. On the other
hand, at the very low temperatures, enhancement of $\lambda$ would
improve the ETE and brings it to a saturated high level. The
gradient and Hessian norms as functions of reorganization energy and
temperature are illustrated in Fig. \ref{Figopt-sus-lamT}. At the
relevant FMO operating temperatures, optimum and robust energy
transport can occur simultaneously only within a small regions of
$\lambda$ between $30$ to $35$ $cm^{-1}$, which coincide with the
estimated values of reorganization energy for the FMO. We note that
there are certain regions of higher robustness at higher
reorganization energy that are in principle available, but these
regions imply a significantly lower operating temperature for the
FMO operation and they have suboptimal ETE in comparison with actual
FMO environmental parameters at the room temperature.

As shown in the previous section, the parameter $\lambda /\gamma$
governs the shape of the ETE landscape at a fixed high-temperature
limit. However, in the perturbative limit the decoherence rate can
be expressed by $\lambda T/\gamma$ which captures the suboptimal ETE
in the weak localization region. Now, we investigate if the ETE
behavior in all regimes can be globally captured by the parameter
$\lambda T/\gamma$, for the given FMO Hamiltonian. Specifically, we
need to verify if one can predict the optimal noise-assisted
transport region as well as ETE suppression levels at the strong
localization regions by a single parameter $\lambda T/\gamma$. To
examine the validity of this theory, we study ETE as a function of
the reorganization energy and the inverse temperature for the fixed
$\gamma=50$ $cm^{-1}$, see Fig. \ref{topT}. Similar to the plot of
efficiency as a function of $\lambda$, $\gamma$, Fig. \ref{topview},
here also the efficiency landscape is divided by lines of
approximately $\lambda T$  (with some deviations from linearity in
the high $\lambda$ and low temperature regime). It can be observed
from the Fig. \ref{topT} that for small $\lambda T$ weak
localization is dominant. At the intermediate $\lambda T$ values
environment-assisted energy transport occurs. As we move towards
larger system-bath interactions and higher temperatures, strong
dynamical disorder diminishes the coherence and the exciton
migration can be fully described by an incoherent hopping process,
since the wave function is essentially localized over spatial sites.
At this regime, the effect of high temperature can be understood
from the dynamics of BChls energy fluctuations which is described by
the symmetrized correlation function
$S(t)=\frac{1}{2}\langle\{\tilde{B}(t),\tilde{B}(0)\}\rangle_{ph}$
($B_j=B$, for any BChl $j$). The function can be extracted
experimentally by three-pulse photon echo peak shift measurement.
For a Lorentzian density, $S(t)=2\hbar \lambda/\beta e^{-\gamma t}$,
and fixed $\gamma$, the temperature and the reorganization energy
determines the amplitude of the site energy fluctuations. Our
simulation in Fig. \ref{Figefflamtemp} demonstrates that a high
efficient energy transfer can be achieved at a \emph{moderate} site
energy fluctuations away from both weak and strong localization
limits.

Combining different regimes of three important environmental
parameters, $\lambda$, $\gamma$, and $T$, the effective decoherence
strength $\lambda T/\gamma$ emerges as the parameter that governs
the energy transfer efficiency landscape. For the FMO free
Hamiltonian energy gaps, by increasing the single parameter $\lambda
T/\gamma$ from small to intermediate, and from intermediate to large
values, one can describe the transition from weak localization to
ENAQT, and from ENAQT to strong localization. More generally, when
the effective decoherence rate $\lambda T/\gamma$ is either much
smaller or much larger than the typical energy splitting $g$ between
delocalized energy eigenstates, then transport is suppressed. Thus,
in order to predict the general patterns of quantum transport in
generic light-harvesting systems, we should compare the relative
strength of $\lambda T/\gamma$ (with dimension of energy) to the
average excitonic energy gap of the free Hamiltonian that can be
quantified as $g=\frac{1}{N-1}||H-Tr(H)\mathbb{I}/N||_*$ for an $N$
level system, where the nuclear norm $||X||_*$ is defined as the
summation of $X$'s singular values. This follows from the fact that
the ETE of a system with a rescaled Hamiltonian $\alpha H$ equals
the ETE of a system with Hamiltonian $H$ for which other
environmental energy/time scales are renormalized by a factor
$1/\alpha $. Here we give a simple proof for this fact:

\begin{figure*}[tp]
\includegraphics[width=18cm,height=6cm]{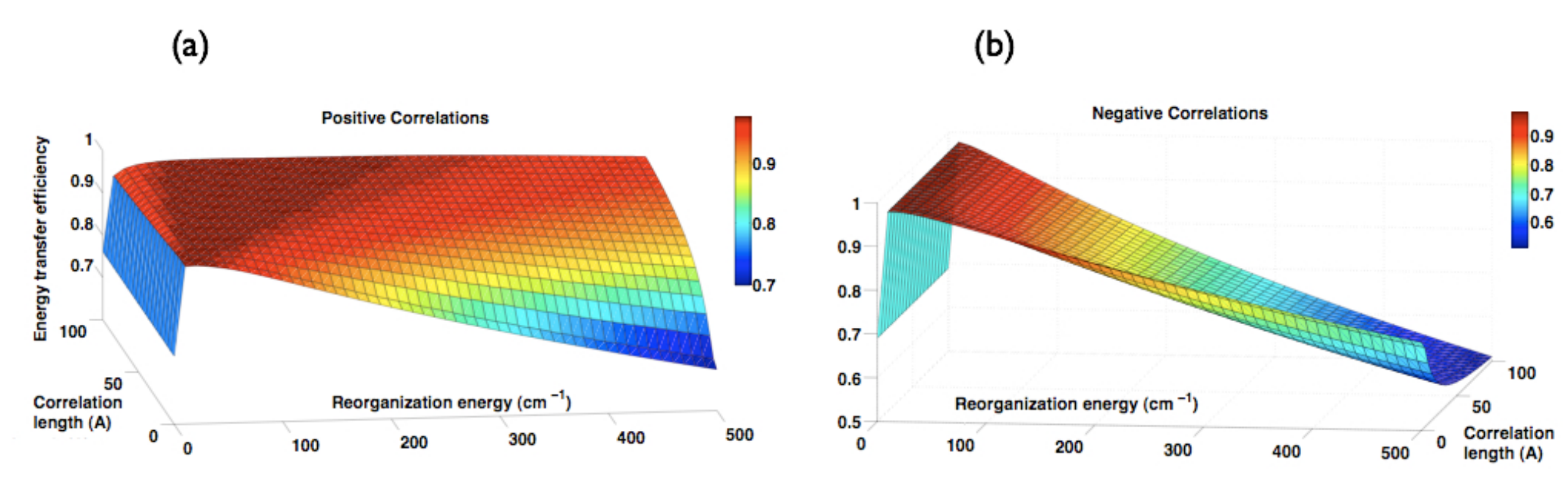}
\caption{(a) ETE as a function of reorganization energy and positive
bath spatial correlations that correspond to in-phase fluctuations
of two spatially separate bath oscillators. At the large values of
reorganization energy, the positive correlations induce certain
symmetries in the effective phonon-exciton Hamiltonian significantly
enhancing ETE, by protecting the system against strong dynamical
disorder similar to decoherence free subspaces \cite{DFS}. In the
intermediate values of $\lambda$, the ENAQT become more robust due
to spatial correlations. In large correlation length, the optimal
ETE is essentially expanded over a much wider regime, one order of
magnitude larger than the optimal ETE region in the absence of any
spatial correlations. At smaller values of $\lambda$ positive
spatial correlations actually hinder the exciton transport by
decoupling useful but very weak fluctuations. A similar linear
relationship is observed here between $\lambda$ and positive spatial
correlations as those with $\gamma$ and $T$. (b) ETE as a function
of reorganization energy and negative spatial correlations that
correspond to out of phase fluctuations of two spatially separate
bath oscillators. Negative correlations reduce ETE at all values of
$\lambda$.} \label{Figeffcor}
\end{figure*}

The efficiency defined in Eq. (\ref{ETE}) can be simply calculated
as $\eta=2r_{trap} \langle trap|\tilde{\rho}(s=0)|trap\rangle$ where
$\sim$ denotes the Laplace transform. The operator
$\tilde{\rho}(s=0)$ can be found by transforming the dynamical
equation (\ref{TNME}):

\begin{eqnarray}
-\rho(0)&=&\mathcal{L}_S\tilde{\rho}(0)+\mathcal{L}_{e-h}\tilde{\rho}(0)\\\notag
&-&\sum_j[S_{j},\frac{1}{\hbar^2}\tilde{C}(-\mathcal{L}_S)
S_{j}\tilde{\rho}(0)-h.c.].
\end{eqnarray}

This yields the ETE
\begin{eqnarray}
\eta=-2r_{trap} \langle trap|\mathcal{K}\rho(0)|trap\rangle
\label{ETE2}
\end{eqnarray}
where
\begin{eqnarray}
\mathcal{K}=(\mathcal{L}_S+\mathcal{L}_{e-h}-\sum_j[S_{j},\frac{1}{\hbar^2}\tilde{C}(-\mathcal{L}_S)
S_{j}-h.c.])^{-1}
\end{eqnarray}

The Laplace transform of the correlation function (\ref{lowT}) can
be found as
$\tilde{C}(-\mathcal{L}_S)=\frac{\gamma\lambda}{\hbar(\gamma-\mathcal{L}_S)}(\cot(\frac{\beta\hbar\gamma}{2})-i)+
\frac{4\lambda\gamma}{\beta\hbar^2}\sum_{k=1}^{\infty}\frac{\nu_k}{(\nu_k^2-\gamma^2)(\nu_k-\mathcal{L}_S)}$.
Considering these explicit expressions, it is easy to see that the
system-bath parameters $\{\alpha
H,\lambda,\gamma,T,r_{trap},r_{loss}\}$ yields the same ETE
(\ref{ETE2}) as the parameter set $\{H,\lambda/\alpha,\gamma/\alpha
,T/\alpha,r_{trap}/\alpha ,r_{loss}/\alpha \}$. We should mention
that similar results can be obtained for the hierarchy equation of
motion.

Overall, we introduce the dimensionless parameter as $\Lambda =
\lambda T/\gamma g$, as the parameter that governs the energy
transfer efficiency landscape. When we approach $\Lambda\approx1$,
the decoherence rate is tuned to give the maximum transport rate.
For $\Lambda \ll 1$ and $\Lambda \gg 1$ we move toward weak and
strong-type of transient localizations respectively. To examine the
temperature and also spectral density independency of the observed
patterns in Figs. \ref{topview} and \ref{topT}, we present the
landscape top view of ETE for a) Lorentzian density at $77^\circ K$
and b) Ohmic density at $298^\circ K$, in appendix B. In both cases
a similar pattern can be observed.

In the subsequent sections, we fully explore the interplay of
reorganization energy with a variety of Frenkel-exciton free
Hamiltonians for systems consist of up to 20 chromophores with
modest site energetic disorders. Our numerical studies suggest that
in the optimal transport regime the relevant energy scale of the
free Hamiltonian, $g$, can be essentially expressed as the average
dipole-dipole interactions strength that is $\left\vert \mu
\right\vert ^{2}n$, where is $\left\vert \mu \right\vert ^{2}$ is
the average dipole-moment strength and $n=\frac{8N}{D^{3}}$ is the
density of $N$ chromophores homogenously embedded in a system with
size $D.$ We will address the significance of the convergence of
energy scales in the broader context of complex quantum systems in
the discussion section.

\subsection{Reorganization energy and bath spatial correlations}

Recently there have been significant interest on the potential role
of bath spatial correlations as the underlying physical principle
leading to the experimental observations of long-lived quantum
coherence in biomolecular systems \cite{Engel07,Lee07,panit10}.
These correlations have also been the center of attention and debate
in the recent theoretical studies of quantum effects in
photosynthetic complexes
\cite{Rebentrost08-1,Olaya10,AkiPCCP10,Sarovar11} as a potential
positive feature of environmental interactions. The basic intuition
is that if the environmental fluctuations are correlated in space,
e.g., in the site basis, then they will lead to fairly global
quantum phase modulations in exciton basis, similar in spirit to
\emph{decoherence-free subspaces} that are studied in details in
quantum information science \cite{DFS}. Indeed, as demonstrated
earlier in Ref. \cite{Rebentrost08-1}, using a Lindblad master
equation, positive spatial correlation can enhance the underlying
contributions of quantum coherence to ETE according to two different
measures based on Green's function methods. However, the overall ETE
remains relatively unchanged for the FMO estimated values, as the
contribution of relaxation to ETE drops with a similar rate Ref.
\cite{Rebentrost08-1}. Here, we are interested in the potential role
of quantum correlations in optimality and/or flexibility of energy
transport rather than enhancing the time-scale of quantum coherence
beating.

\begin{figure*}[tp]
\includegraphics[width=18cm,height=6cm]{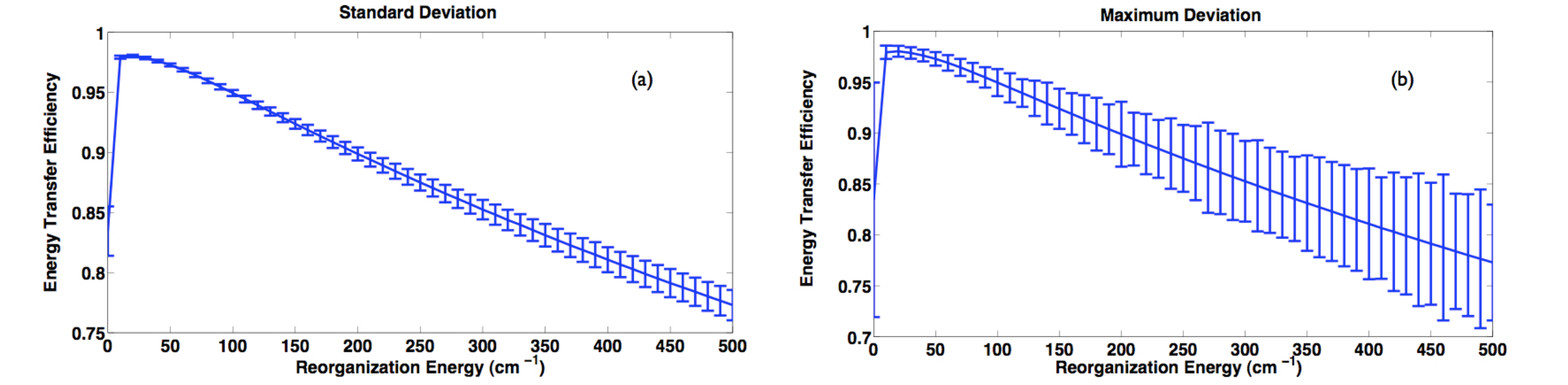}
\caption{The degree of sensitivity of ETE for $10^4$ uniformly
sampled pure and mixed initial exciton density matrices for
different values of reorganization energy: (a) The error bars
indicate the ETE standard deviation. At the FMO value of $\lambda=35
cm^{-1}$ the standard deviation of ETE has a negligible value of
about $0.1\%$. (b) A worst-case scenario of FMO energy transport
sensitivity to initial exciton states. Here, the error bars indicate
the maximum and minimum values of ETE achieved over the sample of
$10^4$ randomly chosen initial states. At the FMO value of
$\lambda=35 cm^{-1}$ the variation in ETE is less than $1\%$}
\label{Figinitial_state_ave}
\end{figure*}

The TC2 can be generalized to include the effect of both spatial and
temporal correlations between protein environments:
\begin{eqnarray}
&&\frac{\partial}{\partial t}\rho(t)=\mathcal{L}_S\rho(t)+
\mathcal{L}_{e-h}\rho(t)
\label{TNMECOR}\\
&&-\sum_{j,k}[S^j,\frac{1}{\hbar^2}\int_0^t C_{j,k}(t-t')
e^{\mathcal{L}_S(t-t')} S^k\rho(t') dt'-h.c.]\notag
\end{eqnarray}
The function $C_{j,j}$ is just the autocorrelation function of the
$j $th environment. The cross correlation between fluctuations of
sites $j $ and $k$ is given by the function $C_{j,k}(t-t')=\langle
\tilde{B}_j(t)\tilde{B}_k(t')\rangle$. In our study we assume
$C_{j,k}$ to be of Lorentzian
$\lambda_{j,k}\gamma\omega/(\gamma^2+\omega^2)$ form with the
cut-off frequency $\gamma$, similar to the autocorrelation spectral
density, and the strength $\lambda_{j,k}$ that decays exponentially
with the distance between Bchls $j$ and $k$, $d_{j,k}$. A
correlation length $R_{cor}$ determines the strength of the spatial
correlations, $\lambda_{j,k}=\lambda \exp(-d_{j,k}/R_{cor}) $. We
distinguish between the two different cases of positive and negative
correlations, corresponding to $\lambda>0$ and $\lambda<0$
respectively.  The sign of the correlations is imposed by the
position and orientation of the Bchls with respect to their
surrounding protein \cite{NPcorrelations}.

The effects of positive and negative spatial correlations on the
energy transfer efficiency for various degrees of the system-bath
couplings are presented in Fig. \ref{Figeffcor}. One main
observation is that both positive or negative spatial correlated
environmental fluctuations play an insignificant role in the overall
ETE only at the regions of reorganization energy very close to
$\lambda=35 cm^{-1}$. However, as we slightly increase the
system-bath coupling strength, we see two very distinct behaviors of
the ETE as a function of long-range spatial characteristics of the
bath fluctuations, depending on whether these variations being
correlated or anti-correlated. For rather strong reorganization
energy, positive spatial correlations can enhance the ETE by about
$30\%$. By contrast, negative spatial correlation can reduce the ETE
by the same amount for reorganization energy of above $\lambda=400
cm^{-1}$.

A careful inspection of Fig. \ref{Figeffcor} (a) shows a linear
relationship between reorganization energy and positive spatial
correlations similar to those with $\gamma$ and $T$, as discussed in
the previous section. For very small reorganization energies, when
quantum localization due to static disorders is present, the
positive spatial correlations can slightly reduce ETE by decoupling
the fragile but positive role of environmental fluctuations. On the
other extreme regime, positive spatial correlations can play a
significant positive role by inducing an effective \emph{symmetry}
in the system-bath interactions, leading to partial immunity with
respect to the strong dynamical disorders at large reorganization
energies. Remarkably, in the intermediate regime, the induced
symmetries can substantially enhance the robustness of the
environment-assisted transport. We can observe that the width of
ENAQT region (in reorganization energy axes) is enhanced from $20
cm^{-1}$ for zero correlation length to $200 cm^{-1}$ for $100 \AA$
correlation length. At this wide regimes, the system is protected
with respect to the adversarial effects of dynamical disorder
preventing the strong localization effects. Therefore, the spatial
correlations should be also incorporated in the governing parameter
$\lambda T/\gamma$. This can be achieved by renormalizing the
reorganization energy $\lambda$ to an effective lower/higher
values, when positive/negative correlations exist. 

In order to quantify the role of spatial correlations in enhancing
the quantum coherence time-scale of light-harvesting complexes, one
has to partition the time-nonlocal master equation, Eq. \ref{TNME},
similar in spirit to those studies in Ref.
\cite{Rebentrost08-1,CaoSilbey}. In the other sections of this work
we assume spatial correlations to be negligible for the FMO complex
consistent with the recent simulations in Ref. \cite{Olbrich11}.
Indeed, the atomistic simulation of the FMO pigment-protein-solvent
dynamics at room temperature presented in Ref \cite{Olbrich11}
reveals insignificant correlations in the site energy fluctuations,
suggesting that the uncorrelated bath approximations are reasonably
valid. Next, we explore the role of initial conditions.

\section{Energy transport sensitivity on the initial excitations}

The exciton migration pathways and time-scales have been studied in
detail for a variety of light-harvesting complexes using various
perturbative techniques including F\"{o}rster models for studying
LHI and II of purple bacteria \cite{Ritz} and Lindblad models for
simulating the dynamics of the FMO protein of green sulphur bacteria
\cite{mohseni-fmo}.
Nevertheless, the role of initial conditions in the overall energy
transfer efficiency of photosynthetic complexes is to a large extent
unknown. It was recently shown that the initial quantum coherence
could influence the energy transfer efficiency in LHI of purple
bacteria assuming no interaction with the phonon bath
\cite{Castro08}. The dependency on initial localized excitation at
BChls 1 and 6 were also examined for the FMO complex using Lindblad,
Haken-Strobl, and HEOM models
\cite{mohseni-fmo,Rebentrost08-2,AkiPNAS}. However, the sensitivity
of ETE with respect to generic initial pure and mixed states taken
from a large ensemble in the single-excitation manifold has not
previously been explored.

\begin{figure}[tp]
\includegraphics[width=9cm,height=6cm]{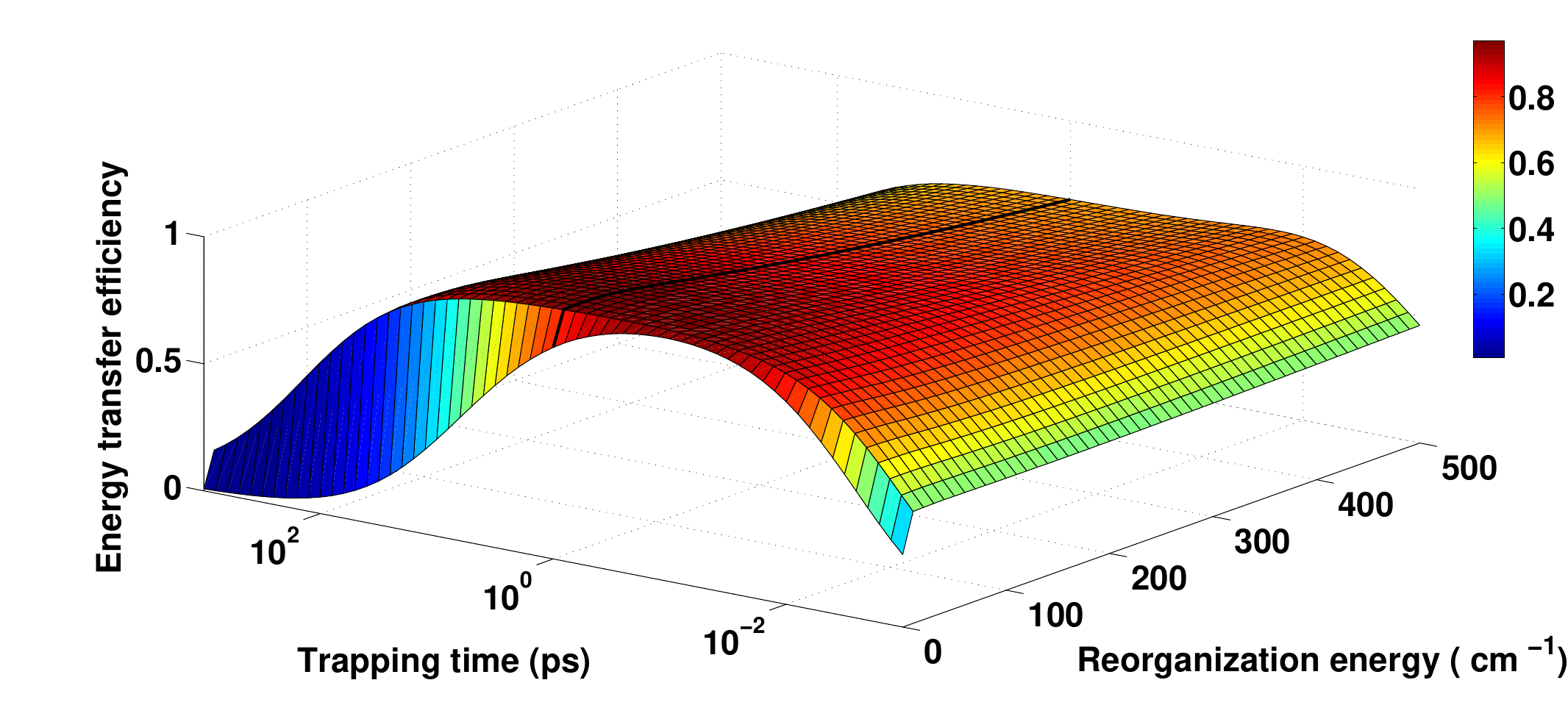}
\caption{The ETE manifold on the parameter space of reorganization
energy and trapping time-scale. It can be seen that the FMO complex
reaches its maximum functionality at trapping rates of about $0.5-5$
$ps^{-1}$. The tunnel-shape ETE landscape can be understood by
noting that at low trapping rates the transport efficiency is
diminished by the recombination process. At high trapping rates the
exciton transfer is suppressed via quantum Zeno effect as the
trapping process corresponds to very strong and continuous
measurement of the system.} \label{Figefftrapcorrect}
\end{figure}

Here, we first examine the average sensitivity of ETE with respect
to randomly chosen initial states for various reorganization
energies. To this end, for each value of reorganization energy, we
sample over $10^4$ (pure or mixed) density matrices from a uniform
distribution in the space of all $7\times 7$ trace one positive
matrices. In Fig.~\ref{Figinitial_state_ave} (a) the average values
of ETE is plotted with an error bar representing the variance of ETE
in our random sampling. Note that at the optimal ETEs, corresponding
to the value of reorganization energy of the FMO complex, the
dependency of the variances on initial states is very small -- less
than $0.1\%$. However, the ETE fluctuations can grow by an order of
magnitude for larger or smaller values of $\lambda$. We also
investigate the best and worst possible random initial single
excitonic states in the Hilbert space of FMO. In
Fig.~\ref{Figinitial_state_ave} (b), we illustrate this extreme
possible deviations by error bars on the average ETE at any given
value of $\lambda$. Note that ETE is very robust, varying about
$1\%$ with respect to different initial excitations at the optimal
area of ETE landscape. However, this robustness diminishes
substantially at the regimes of large reorganization energy. Next,
we study the ETE landscape as a function of trapping and dissipation
rates.

\section{Temporal and geometrical effects of the trapping mechanism}

\begin{figure}[tp]
\includegraphics[width=9cm,height=6cm]{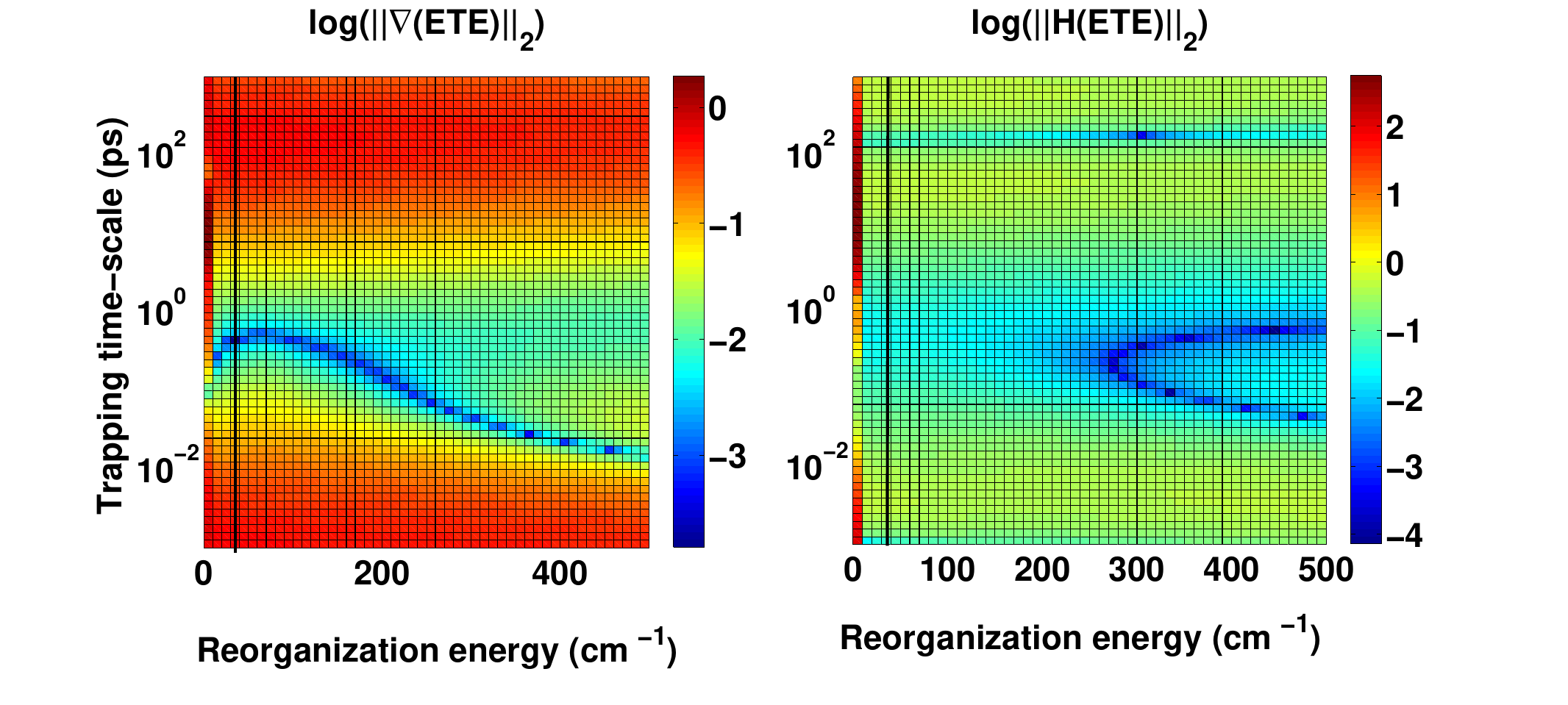}
\caption{This figure is a complement to Fig. \ref{Figefftrapcorrect}
(Left). The degree of ETE optimality is quantified at different
values of $\lambda$ and $r_{trap}^{-1}$ by the norm of the ETE
gradient. The dark blue points represent near optimal values.
(Right) The degree of ETE robustness is quantified by the Euclidian
norm of the ETE Hessian. The blue points represent near robust
points. The FMO achieves its maximum efficiency at
$r_{trap}^{-1}=0.5 ps$. Note that for larger reorganization energies
the trapping rate has be increased to achieve optimal ETE. However,
this competition does not exist at small and intermediate
system-bath coupling strength, where $\lambda$ is on the order of
off-diagonal elements of the FMO free Hamiltonian leading to
environment-assisted energy transport. This suggests that a general
convergence of time-scale might be required to obtain global
efficient and robust transport (see discussion section). Although
certain regions of highly robust ETE can be observed in figure (b),
but those cases are suboptimal by considering their corresponding
values in figure (a).} \label{Figopt-sus-lamtrap}
\end{figure}

Basic structural information on the FMO-RC complex has been obtained
via linear dichroism spectra and electron microscopy
\cite{Remigy99}. These studies indicate that the symmetry axis of
the trimer is normal to the membrane containing the reaction center.
The electron microscopy resolution is generally not sufficient to
distinguish the top and the bottom of the FMO trimer nor the
distance between FMO-RC. Thus, in principle either pairs of BChls 1
and 6 or BChls 3 and 4 are the pigments that connect the FMO complex
to RC. However, it is widely believed, due to efficient energy
funneling toward RC, that BChl 1 and 6 are the linkers to antenna
baseplate, and 3 and 4 should serve as target regions within the
neghibourhood of RC complex \cite{Renger06}. This hypothesis has
been recently verified experimentally \cite{Wen09}.  Up to this
point, we have considered BChl 3 to be in the close proximity of RC
by a trapping time-scale of about $1$ $ps$.  However, in this
section we consider both of these parameters to be free, in order to
explore the optimality and robustness of the ETE landscape as we
vary the time-scale and geometrical constraints set by the RC
trapping mechanism.

In Fig. \ref{Figefftrapcorrect}, we study the behavior of energy
transfer efficiency landscape in various trapping time-scales and
reorganization energies. It is evident that as the trapping rate
becomes very slow comparable to $100$ $ps$ or slower, the ETE drops
significantly independent of the values of $\lambda$. This can be
understood intuitively as follows: the excitation has to wait on
average so long for successful trapping to take place such that
there will be an increasing chance of electron-hole recombination as
we are approaching time-scales comparable to exciton life time. If
the trapping mechanism occurs within a time-scale of $1$ $ps$, the
ETE reaches to its expected maximum value of about 99\%. Generally,
one might expect that with increasingly faster trapping mechanisms
the likelihood of dissipation to environment vanishes and energy
transport approaches to the ideal case of having perfect efficiency.
However, when the trapping rate becomes very fast on the order of
$10^{-2}ps$ or faster, the ETE also drops significantly, a result
that might appear counter-intuitive.   In fact, overly rapid
trapping leads to low efficiency via the quantum Zeno effect, as the
rapid trapping effectively freezes the exciton dynamics and prevents
it from entering the reaction center. As a result, the finite
exciton life-time eventually leads to complete dissipation of
excitation to the environment in extreme limit of fast trapping of
about $1fs$.

\begin{figure}[tp]
\includegraphics[width=9cm,height=6cm]{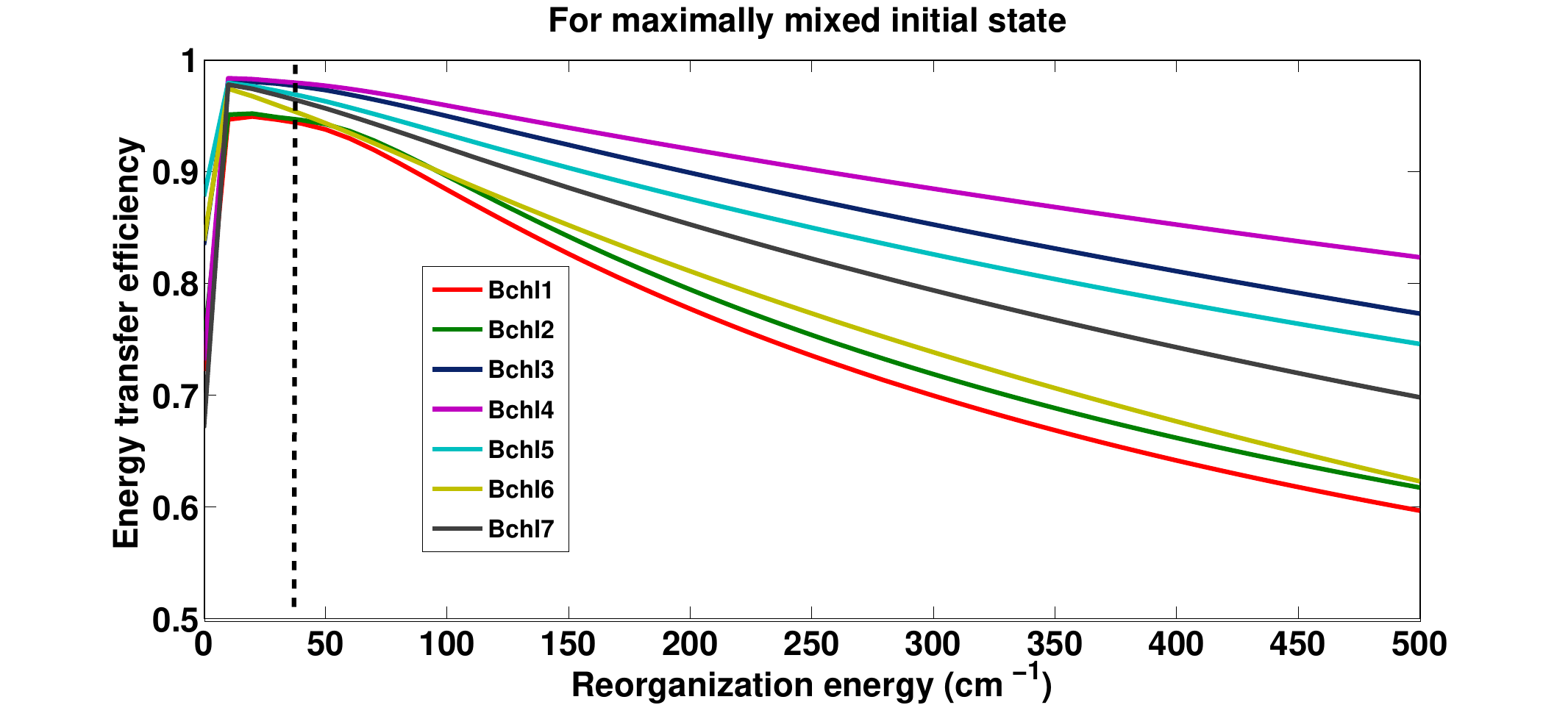}
\caption{The ETE as a function of reorganization energy for the
initially maximally mixed state. In each plot one of the 7 BChls is
considered to be connected to the reaction center. BChls 3 and 4
acting as the exciton transfer bridge yield the highest efficiency
for almost all values of the reorganization energy. This confirms
the experimental evidence that the FMO spatial orientation is such
that the BChls 3 and 4 are located near the RC.} \label{Figsites}
\end{figure}

The optimality and robustness of ETE versus both decoherence and
trapping rates using gradient and Hessian norms are presented in
Fig. \ref{Figopt-sus-lamtrap} (left panel). It can be observed that
at $\lambda=35 cm^{-1}$ for the FMO, the ETE is optimal with a
trapping rate of about 0.5ps. If the environmental interactions were
stronger, a comparably faster trapping mechanism would be required
to preserve such high level of efficiency.  However, for small and
intermediate system-bath interaction strength, where
environment-assisted transport occurs, slower trapping rates become
optimal, that is $\lambda$ and $r_{trap}$ are not competing
processes anymore. This implies that a multi-parameter convergence
of time-scales of the relevant physical processes might be required
for light-harvesting complexes to operate optimally. We will address
this issue in a broader context in the discussion section. From Fig.
\ref{Figefftrapcorrect}, it can be easily seen that ETE is very
robust to variation of trapping rate at about $1$ $ps$ time scale.
In Fig. \ref{Figopt-sus-lamtrap}, we also illustrate the robustness
with respect to both trapping and reorganization energy (right
panel). For rather large values of $\lambda$, there are certain
regions that are highly robust to both parameters, but they are in
fact suboptimal, as can be seen from noting their values in the left
panel.

\begin{figure}[tp]
\includegraphics[width=9cm,height=6cm]{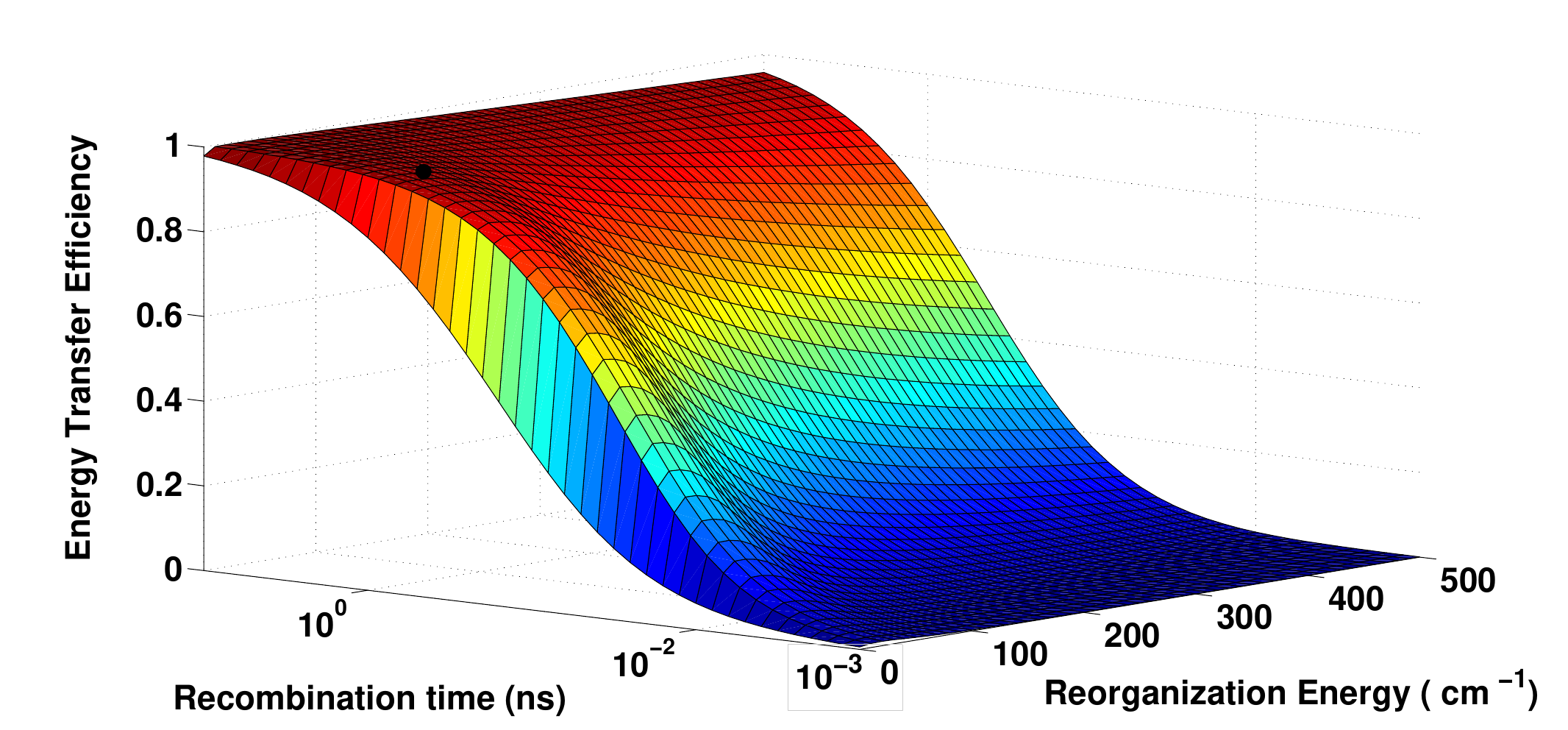}
\caption{ETE versus dissipation (loss) time-scale $r_{loss}^{-1}$
and reorganization energy. The maximum optimality and robustness for
FMO is observed around the estimated value $r_{loss}^{-1}=1ns$
implying the significance of the time-scale separation between
dissipation and trapping rates. We note that the ENAQT effect is
ubiquitous at all rates of electron-hole recombination process.}
\label{efflamrecomb}
\end{figure}

To explore the optimality of ETE with respect to the location of
reaction center, we consider the efficiency of other scenarios that
the reaction center can be in the proximity of any other BChl sites.
Figure \ref{Figsites} shows a cross-section of Fig.
\ref{Figeffgamlam}, in a fixed $\gamma=50 cm^{-1}$, for all possible
trapping sites. To be unbiased with the respect to the initial
state, we assume a maximally mixed initial state. It can be seen
that the optimal curves belong to BChls 3 and 4 as expected, since
they contribute highly to the lowest energy delocalized excitonic
states. It should be noted that optimal environment-assisted quantum
transport, and the two extreme regimes of quantum localization can
be observed for all of these plots independent of the actual
location of trapping.  In other words, the behavior of the energy
transport efficiency landscape and its dependence on a single
governing parameter are not properties of a particular choice of
trapping site in the FMO structure.

For completeness, we also investigate the ETE landscape as a
function of dissipation (loss) rate and reorganization energy in
Fig. \ref{efflamrecomb}. In our simulation of the FMO dynamics we
have used the estimated value of $r_{loss}^{-1}=1ns$. In Fig.
\ref{efflamrecomb}, however,  we treat loss rate as a free parameter
and we observe that for any stronger dissipation process, ETE would
have a suboptimal and less stable behavior. Thus, even if all the
other important parameters are within the optimal regime, a large
time-scale separation between dissipation and trapping rate is still
required to guarantee the highest performance for light-harvesting
complexes. Fig. \ref{efflamrecomb} also demonstrates that the
existence of ENAQT is independent of a particular choice of
dissipation rate.

\begin{figure}[tp]
\includegraphics[width=9cm,height=6cm]{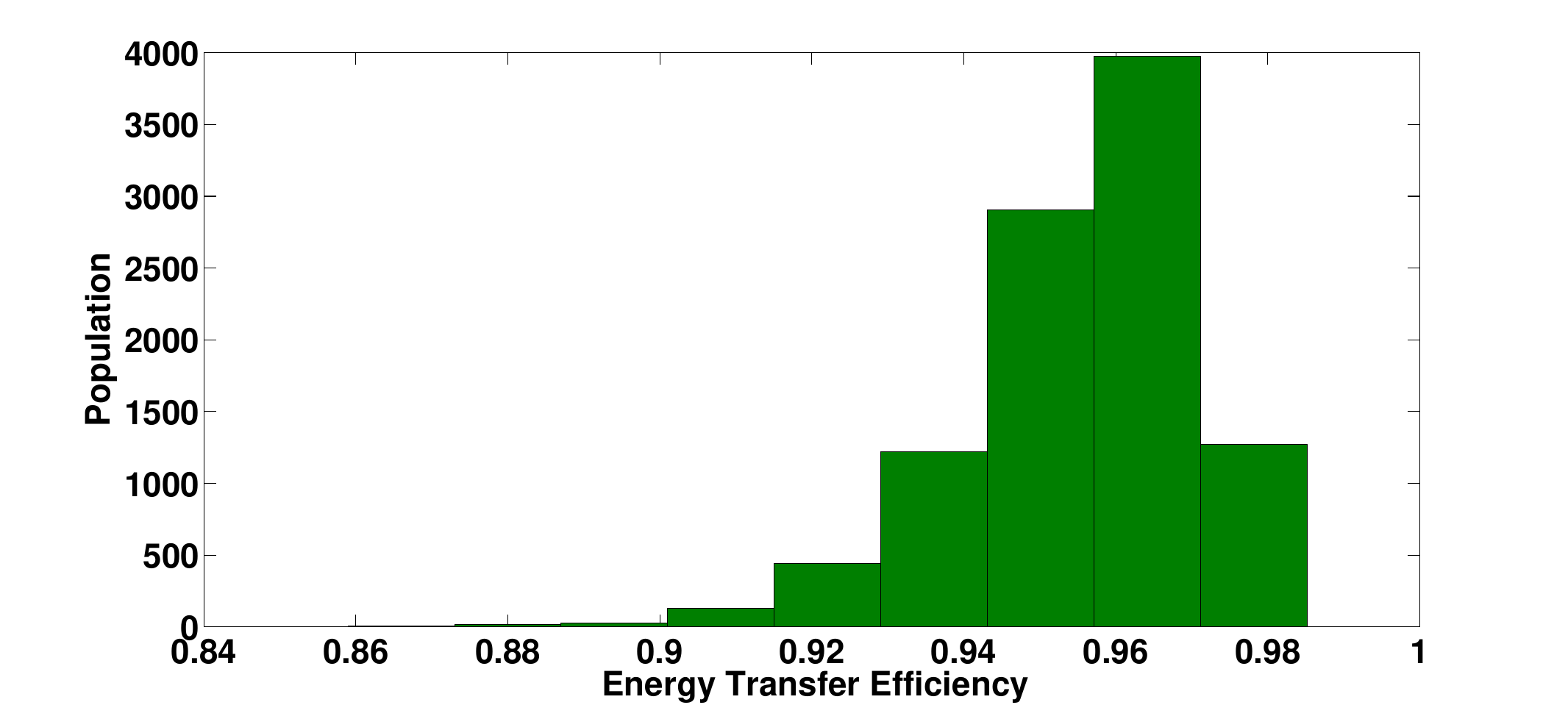}
\caption{Robustness of FMO transport efficiency with respect to
small variations of BChls locations, site energies and dipole
orientations for 10000 samples: the Hamiltonian parameters are
perturbed with site energies disorders of $\pm 10 cm^{-1}$,
dipole-moment uncertainties of $\pm 5^\circ$, and BChls spatial
displacement of $\pm 2.5 \AA$. The statistical distribution of
$10^4$ random configurations shows a significant degree of
robustness such that $99\%$ of samples still preserve an efficiency
of above $0.9$.} \label{Figconservative}
\end{figure}

\section{optimality and robustness with respect to parameters of FMO
Hamiltonian}

So far, we have demonstrated that for the estimated Hamiltonian
elements of the FMO complex, the environmental parameters and
trapping rates are within the right set of values leading to an
optimal noise-assisted energy transfer efficiency. Moreover, the
performance of FMO is robust with respect to variations in such
decohering and lossy processes and to uncertainties in initial
conditions.  However, in the context of exploring the role of
coherent system evolution, it is not fully clear if the FMO internal
Hamiltonian parameters have evolved to function optimally and fault
tolerantly, despite disorders and thermal fluctuations.  This issue
has been examined for LHCII in Ref. \cite{Sener02} using
semi-classical Pauli master equations to simulate the exciton
dynamics. Here, we would like to use TC2 to explore how rare is the
FMO geometry in terms of its efficiency, whether the specific
spatial and dipole moment arrangements of BChls are essential for
such highly efficient functioning of this pigment-protein complex,
and how robust these parameters are with respect to small and large
perturbations in chromophoric distances, dipole moment orientations,
and site energy fluctuations. Specifically, we explore if the FMO
closely packed structure plays any functional role, and illustrate a
potentially important convergence of the relevant dynamical
time-scales in the FMO energy transport. In the following section,
we investigate the underlying physical principle of quantum
transport in more generic multichromophoric structures beyond the
FMO geometry.  We examine the efficiency of small light-harvesting
complexes as functions of the compactness level, number of sites,
reorganization energy, exciton energy gaps, and multichromophoric
connectivity.

The Frenkel exciton Hamiltonian for a multichromophoric system is expressed as:
\begin{eqnarray}
H_{S} &=&\sum_{j,k}\epsilon _{j}|j\rangle \langle
j|+J_{j,k}|j\rangle
\langle k|,
\end{eqnarray}
in which $J_{jk}$ are Coulomb couplings of the
transition densities of the chromophores,
\begin{eqnarray}
J_{jk}\sim \frac{1}{R_{jk}^{3}}(\mathbf{\mu }_{j}\cdot \mathbf{\mu }%
_{k}-\frac{3}{R_{jk}^{2}}(\mathbf{\mu }_{j}\cdot \mathbf{R}_{jk})(\mathbf{%
\mu }_{k}\cdot \mathbf{R}_{jk})),
\end{eqnarray}
where $\mathbf{R}_{jk}$ denotes the distance between site $j$ and
$k$, and $\mathbf{\mu }_{j}$ is the transition dipole moment of
chromophore $j$ \cite{Damjanovi97}. We first study the robustness of
free Hamiltonian parameters within the proximity of the estimated
values for FMO as given in the appendix A.

\begin{figure}[tp]
\includegraphics[width=9cm,height=6cm]{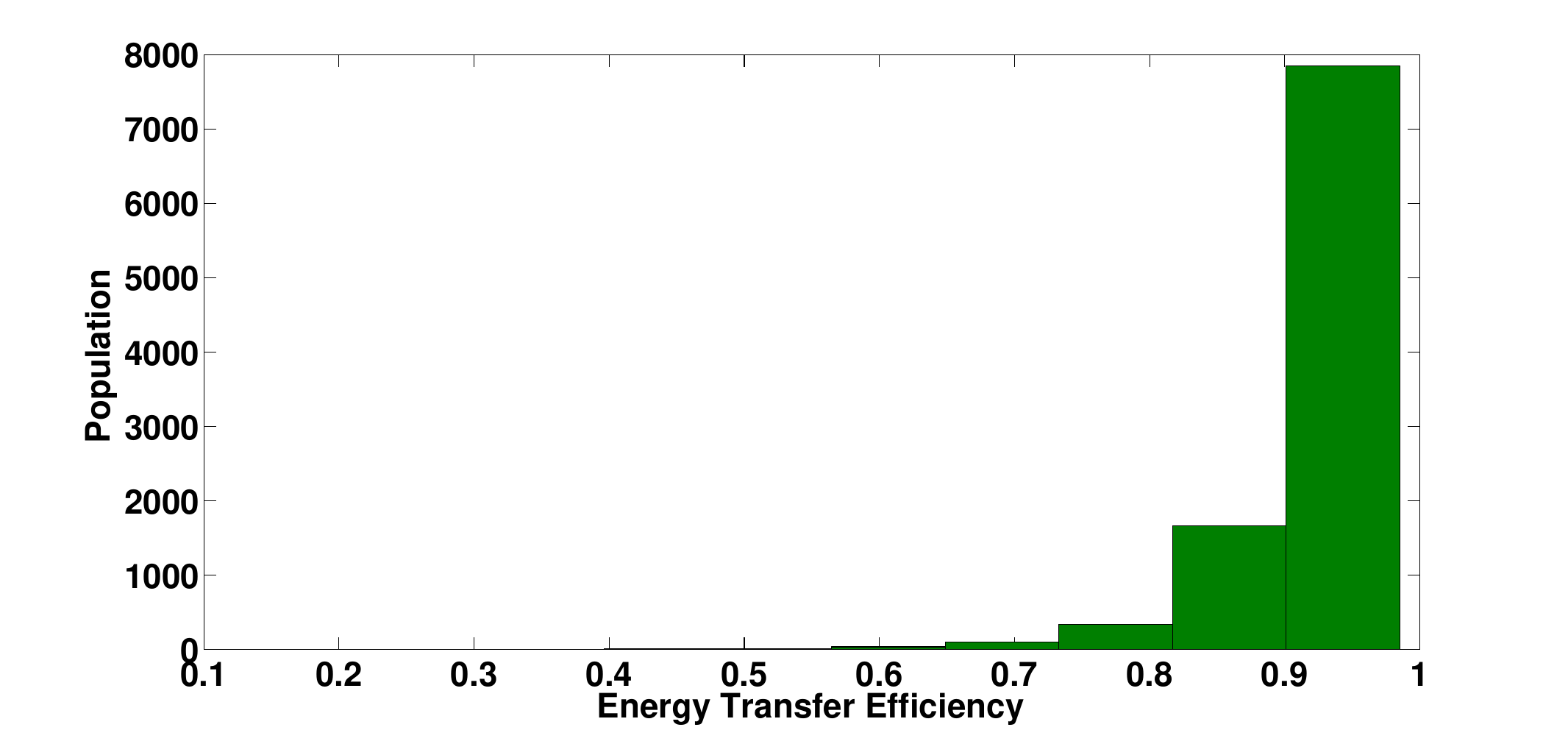}
\caption{Robustness of the FMO complex transport efficiency with
respect to large variations in BChls site energies and dipole moment
orientations for 10000 sample configurations: While the location of
BChls are still slightly perturbed, similar to Fig.
\ref{Figconservative} of about $\pm 2.5 \AA$, the dipole moments can
take any arbitrary direction and site energy takes any value between
zero and $500 cm^{-1}$. This Histogram reveals that the relative
distance of BChls is playing a crucial role in performance of these
random light-harvesting complexes since $79\%$ of them still hold
ETE larger than $90 \%$. } \label{Figrandom_angle-site}
\end{figure}

\subsection{Robustness of FMO Hamiltonian}

Figures \ref {Figconservative} and \ref{Figrandom_angle-site}
demonstrate the robustness of the FMO structure to variations in its
internal parameters. Figure \ref{Figconservative} illustrates that FMO
efficiency does not drop drastically with respect to variations
in the dipoles orientations, site energies, and Bchls
distances close to the neighborhood of the estimated values.
Specifically, from $10000$ random samples of FMO with spatial
uncertainty around each Bchl location of about $\pm 2.5$ $\AA$, dipole
moments orientations variations of $\pm 5^\circ$, and site energy
static disorder $\pm 10$ $cm^{-1}$, $97\%$ of configurations have
efficiency of $95\%$ or higher. This demonstrates a significant
degree of robustness with small perturbations. In order to separate
the influence of spatial coordinates from angular dipole
orientations and disorders, we allow the latter two parameters to
take arbitrary values from a large range while keeping Bchl locations uncertainties
to be limited by $\pm 2.5$ $\AA$. We observe in Fig.
\ref{Figrandom_angle-site} that ETE remains relatively robust with
$79\%$ of random $10000$ configurations have still efficiency of
$90\%$ or higher. This is rather counterintuitive considering huge
freedom that we have accommodated in the dipole moment arrangements
and site energies. These results indicate that spatial degrees of
freedom is a dominating geometrical ingredient of the FMO structure
and might play a key physical role in its performance.

\subsection{Searching for FMO-type Hamiltonians}

To explore the role of the free Hamiltonian parameters, we
first investigated energy transfer efficiencies of 200,000 random
$7$-site chromophoric configurations embedded within a sphere of
diameter $200 \AA$, with random initial and target sites, and with
environmental parameters the same as the estimated values of FMO
\cite{FMOpara}.
We bounded the nearest distance among
chromophores to be larger than $5 \AA$, since the dipole-dipole
approximation breaks down below this limit \cite{Scholes03}. The
dipole moment of each chromophore can take any orientation and the
site energy of each chromophore can take any value between zero and $500
cm^{-1}$ chosen from a random distribution. We observed that only
$10\%$ of these configurations have efficiency larger than $50\%$
and only $0.1\%$ have efficiency comparable to FMO. Thus, one might conclude
that
the FMO structure is one of the rare geometries that can support
this extremely high efficiency. However, as we argue below this conclusion is
inaccurate.

We searched over various geometrical pattern correlations among the
top $0.1\%$ high-end efficient random structures to reveal the
underlying principle(s) that could give rise to their high
efficiency.  We found that the distance between initial
and target sites in most of these structures vary between $6-10 \AA$.
Thus most of these complexes are rather trivial optimal solutions
that are not particulary interesting for quantum transport studies.
Indeed each of those conformations essentially acts as a dimer
including a donor and an acceptor chromophore that are strongly coupled via
dipole-dipole interaction in the near field.  This approach for
optimality is obviously not usable for quantum transport in large
natural or engineered light-harvesting antenna or macromolecular
systems with typically significant spatial separation of initial
excitation and trapping sites.
In order to distinguish potential non-trivial
optimal solutions hidden in these results; i.e., those
involve quantum/classical jumps in a multi-path molecular networks,
we consider the distance of donor and acceptor to be bounded by the
size of the complex for the rest of our studies in this paper.

\begin{figure}[tp]
\includegraphics[width=9cm,height=6cm]{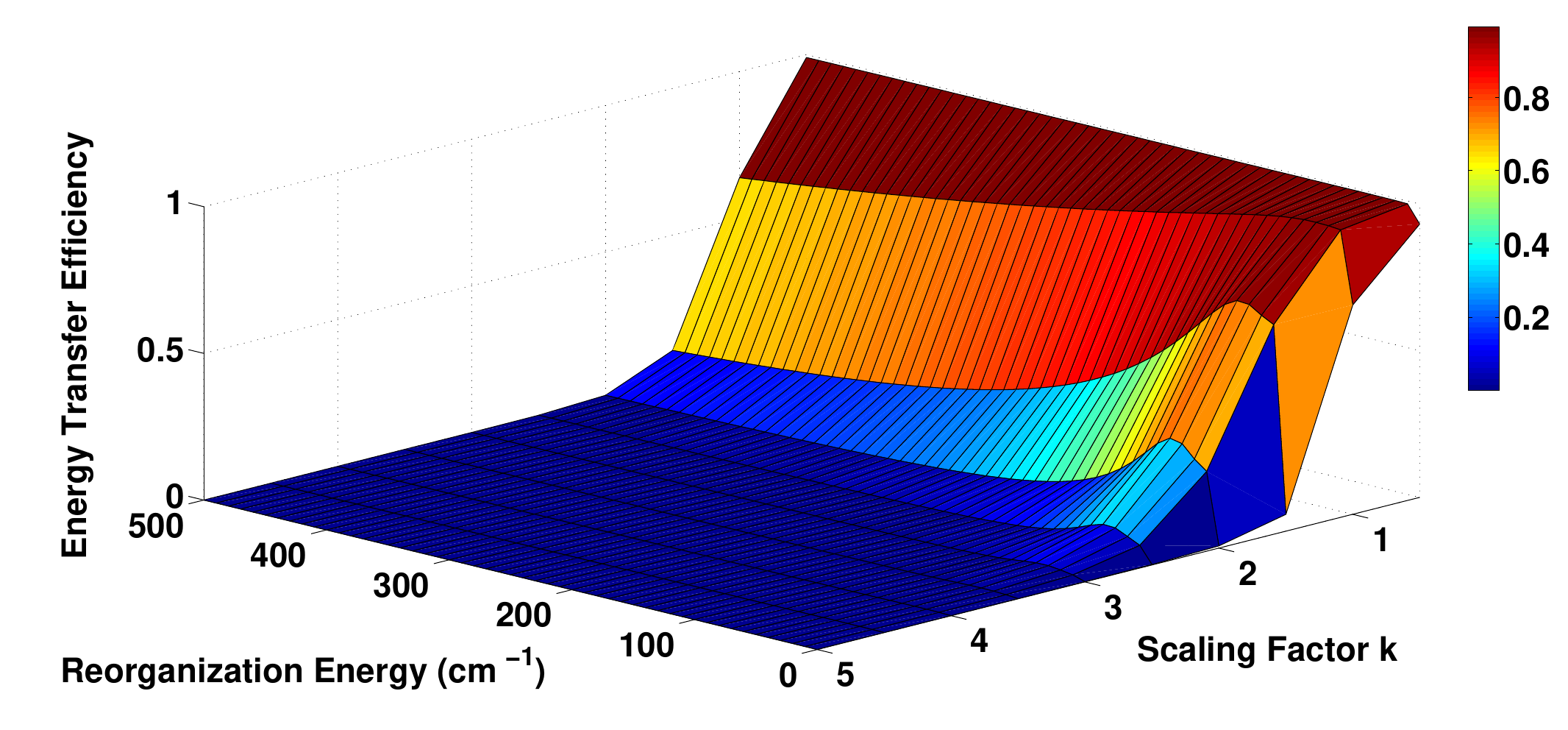}
\caption{Dependency of ETE on compactness level of the FMO complex.
The FMO chromophoric spatial structure is scaled with a factor
between $0.5$ and $5$. The ETE manifold is plotted as a function of
the scaling factor and reorganization energy. For all levels of
compactness, $\lambda=30-40 cm^{-1}$ yields the highest ETE. A more
compact complex shows a higher degree of robustness with the respect
to variation in reorganization energy. The scaling factor $1$
corresponds to the FMO complex leading to a nearly optimal energy
transfer efficiency.} \label{FigcompactnessFMO}
\end{figure}

Our results presented in Figs. \ref{Figconservative} and
\ref{Figrandom_angle-site} clearly indicate that the relative Bchl
locations play a major role in the overall performance of the FMO
complex. Thus a potentially significant parameter of relevance is the
compactness of a given pigment-protein complex. We further explore
this feature by introducing a single \emph{compactness} parameter by
rescaling the Bchl distances by a factor k. We plot ETE as a
function of compactness level k varying by an order of magnitude
from $0.5$ to $5$. To explore any potential interplay of
environmental interactions with this particular internal degree of
freedom, we also simulate this size dependent
supersssion/enhancement of ETE in various reorganization energy,
$\lambda$, in Fig. \ref{FigcompactnessFMO}. It can be seen that
transport efficiency drops significantly by expanding the FMO
structure and has slightly enhanced value for contraction ratio of
$0.5$.  As above, in the discussion of the evolutionary
tuning of reorganization energies and environmental correlation
times, it appears that nature has taken a rather minimal
approach in designing FMO.
By utilizing
the right density level of chromophoric structure and strength of
environmental interactions,
the evolutionary process
has created a nearly optimal energy transfer channel,
but that process apparently avoided the extra work necessary for
further compression to achieve just $1\%$ or less enhancement in
ETE. By inspection of the ETE variation with resect to $\lambda$ in
Fig. \ref{FigcompactnessFMO}, one observes that the ENAQT
phenomenon is scale invariant for the FMO-like structure.

The above analysis reveals an important effective physical
parameter; however, one cannot rule out the possibility of other
underlying dynamical processes and/or geometrical correlations that
could also lead to higher ETE in very different compactness regimes.
In other words, the compactness appears to be a sufficient condition
for high ETE in multichromophoric complexes but most likely it is
not a necessary condition.

\subsection{Convergence of time-scales for FMO}

\begin{figure}[tp]
\includegraphics[width=9cm,height=8cm]{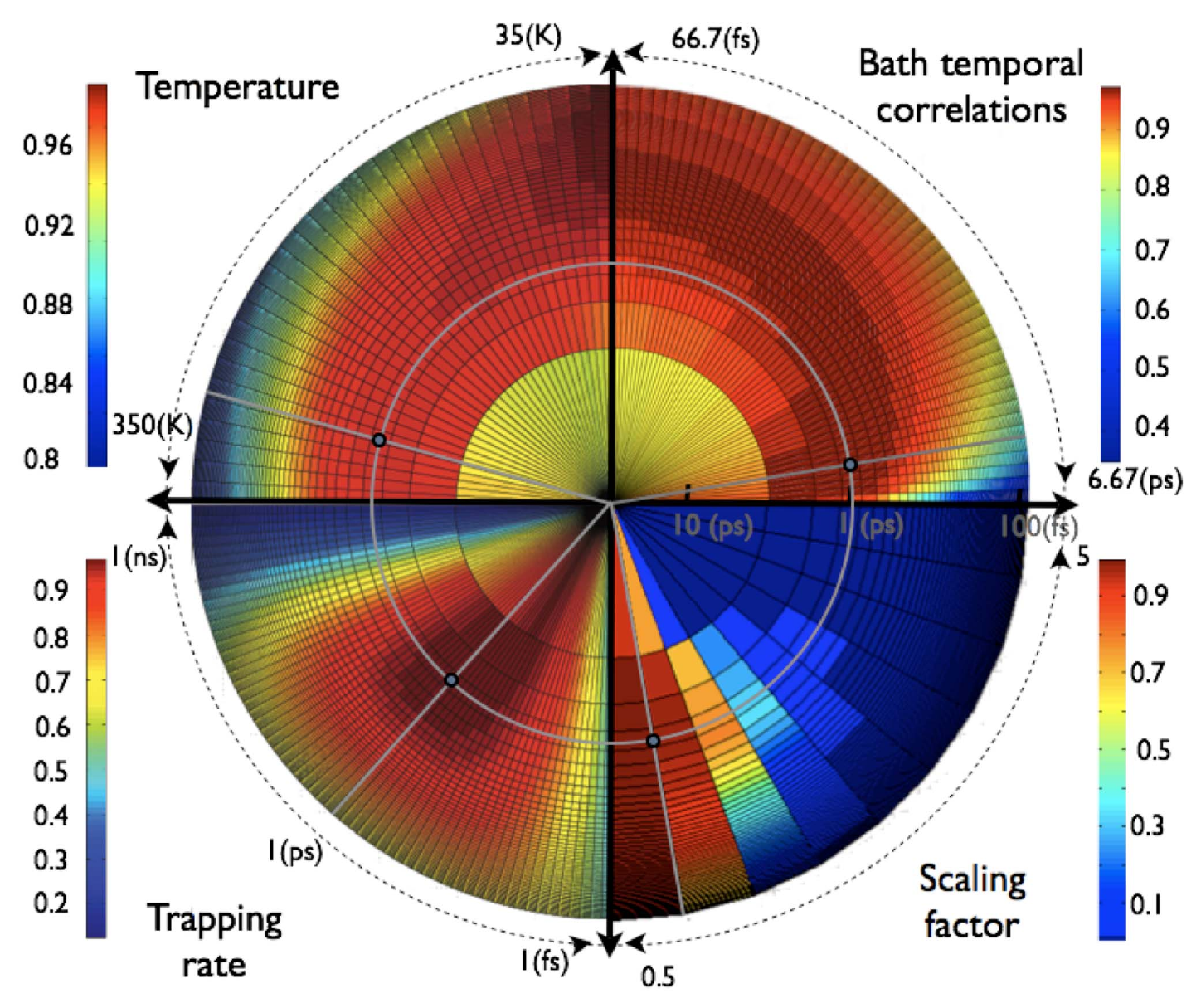}
\caption{A summary of our main results on the optimality and
robustness of FMO energy transfer dynamics illustrating a
convergence of time scale around $1ps$. The radial axes indicate the
reorganization energy in logarithmic scale which is essentially the
main environmental parameter quantifying the strength of system-bath
interactions. Angular coordinate in each quarter represents a
different physical quantity. The first three quarters are associated
to three other important environmental parameters including bath
correlation time-scale, temperature, and trapping time-scale. The
last region highlights the role of compactness as the most
significant internal parameter of FMO structure quantified by a
rescaling factor in the chromophoric relative distances. To improve
the clarity the color bar is rescaled in each quarter. The
estimated values of the FMO complex coincide in the intermediate
values of both angular and radial coordinates, denoted by white
lines (around $1ps$), leading to optimal and robust ETE.}
\label{Figconvergance-timescales}
\end{figure}

We illustrate a summary of our main results on the FMO complex
energy transport dynamics in the non-Markovian and non-perturbative
regimes in Fig. \ref{Figconvergance-timescales}. This figure
indicates that an effective convergence of time-scales for various
dynamical processes exists at the intermediate and optimal regimes
(denoted by a white circle) in the energy transfer landscape
represented by a circular surface. The radius of this circular area
illustrates the reorganization energy in logarithmic scale in the
unit of time from $33.3 ps$ (1 $cm^{-1}$) to $66.7 fs$ ($500
cm^{-1}$). The circular area consists of four different regions in
four quarters, the angular degrees of freedom in each quarter is
associated to a different physical quantity. The first quarter
represents the degree of the memory in the bath or the
non-Markoviaity of the environment in logarithmic scale, that is
equivalent to bath correlation time, varying from about $6.67 ps$ to
$66.7 fs$. The second quarter illustrates variations in the bath
temperature from $35$ to $350^\circ K$. The third region shows six
order of magnitude trapping time-scales from $1ns$ to $1 ps$ in a
logarithmic scale. The last quarter demonstrates compactness level
of the Frenkel exciton Hamiltonian of the FMO rescaled by a factor
from $0.5$ to $5$. The estimated parameter of the FMO is denoted by
a white bar in each region. In all quarters the efficient
neighborhood of ETE are somewhere in the medium range, denoted by a
white circle at about $1$ $ps$ time-scale corresponds to
reorganization energy of the FMO. The optimal areas are far from
extreme low and high reorganization energies in the center and
perimeter of the black circle respectively where quantum
localization effects can substantially suppress ETE.  The key point
of this figure is that all the relevant useful transport time-scales
converge just around $1ps$, far from the dissipation time-scale of
$1ns$. We will discuss the implications of these results in a
broader context of optimality and robustness of complex quantum
systems in the next section.

\section{Discussions on the convergence of timescales: Quantum Goldilocks Principle}

Our numerically study indicates optimality and robustness of FMO
with respect to a variety of parameters. For FMO, in all cases, we
find that the estimated operating values sit nearly at the optimal
point for energy transport efficiency, and that there is a
relatively wide range of parameters about that point for which FMO
is close to optimal. Moreover, we note that the estimated
time/energy scales associated with this point -- couplings between
sites, reorganization energies, decoherence times, environmental
correlation time, and trapping time-scale-- all lie in the same
intermediate range, on the order of a picosecond. Nevertheless it is
instrumental for the FMO performance that life-time of exciton has a
sufficiently longer time-scale. For alternative configurations,
beyond the specific FMO model, we observe that their chromophoric
density, geometrical arrangements, and coupling strengths need to be
in tune with environmental interactions to lead to optimal
performance. Here, we briefly discuss a design principle that
fundamentally relies on this convergence of time scales.

A common design technique in engineering is to keep the desired
system as simple as possible. The rationale is that overly complex
designs are typically not robust: the more complex the design, the
harder it is to predict its behavior, and the more pieces there are
to fail. Now, FMO and other photosynthetic complexes are definitely
not simple: they consist of multiple parts, and, as our studies
show, enlist a wide variety of quantum effects, tuning those effects
to attain efficient and robust operation. Apparently, these
complexes do not obey the fundamental engineering philosophy of
\emph{keep it simple}. There is, however, a more sophisticated
version of the design principle in system engineering, as overly
simple designs may be robust, but they typically also fail to attain
their desired functionality. This advanced version of the design
principle declares that to the extent that it is possible to add
more features to a design to enhance functionality without reducing
robustness, it is desirable to do so. That is, given available
technologies, there is typically a design that attains a level of
complexity that is `just right' -- not so simple that it fails to
attain its goal, but not so complex that it loses robustnesses.  The
design strategy that aims for the just-right level of complexity is
sometimes called the \emph{Goldilocks principle} for complex systems
\cite{Lloyd90}. Let's look at how this principle applies in the case
of the structure of FMO, a complex quantum system that has been
evolved by natural selection.

The Goldilocks principle for complex quantum systems -- the
\emph{quantum} Goldilocks principle-- states that to improve
efficiency, quantum effects could be used, as long as by doing so it
does not make the operation of the system more fragile. Our
simulations show that this is the case with FMO: a wide variety of
parameters have been tuned to their optimal value, and that optimal
value is robust against variations in those parameters. The quantum
Goldilocks principle essentially explains why it is beneficial for
certain parameters that involve time or energy to converge to the
same time/energy scale.

When the time scale for one effect is very different from the time
scale for another, then the first effect can typically be regarded
as a perturbation on the other.  For example, if the environmental
interactions are much slower than the system internal dynamics,
and/or if the correlation time of the environment is much shorter
than the coherent tunneling rate of an exciton from chromophore to
chromophore, then the effect of the environment can be treated in
the perturbative and/or Markovian limits. If one effect is to
interact strongly with another effect, by contrast, then the
time/energy scales of the two effects should be similar.  As we have
shown here, the maximum efficiency of quantum transport is attained
when the key decoherence parameter $\lambda T/\gamma$ is on the same
order as the free Hamiltonian coherent time-scale $g$.  As shown
elsewhere \cite{mohseni-fmo,Rebentrost08-2,Plenio09,Jianlan10}
environmentally assisted quantum transport is an important effect.
This {\it convergence} of time scales is a nuisance for quantum
simulation, as perturbative methods break down.  While this
phenomenon makes the efficient simulation of such systems very
difficult, the convergence of time scales is quite useful for
nature, however, as the two effects can now interact strongly with
each other to produce a significant enhancement in efficiency. Of
course, too strong interaction could also produce a significant
decrease in efficiency. But for systems undergoing natural
selection, one typically expect that the strong interaction is tuned
at the right level to give rise to a beneficial effect.

The results of this manuscript illustrates the quantum Goldilocks
principle manifestation in the FMO complex and other small-size
mutlichromophoric configurations: the time/energy scales have
converged within the range where the different quantum effects and
structural parameters at work can interact strongly with each other
to give rise to possibility of high efficient energy transport.
Moreover, this variety of quantum effects can be combined to attain
high efficiency without sacrificing robustness. The convergence of
time scales makes these complexes difficult to model, but highly
efficient in certain instances.  We anticipate that a similar
convergence of time/energy scales could appear in larger
biomolecular complexes and in biological processes where quantum
mechanics might play an important role.

\section{Conclusion}

In this work, we studied the structural and dynamical design
principles of excitonic energy transfer in the Fenna-Mattews-Olson
(FMO) complex. Our numerical simulations demonstrate that the
natural structure of FMO pigment-protein complex lead to a highly
efficient and robust photosynthetic energy transfer wire. We
characterized the ETE landscape by three main regions: weal
localization, environment-assisted quantum transport, and strong
localization, and identified the scalar $\Lambda =\frac{\lambda
T}{\gamma g}$ as the key parameter to cross between these regions as
one hikes over the landscape.

We showed optimality and robustness of energy transport efficiency
in FMO with respect to variations in all the main external
parameters dictating the dynamics. In particular, we explored the
protein-solvent environment factors: system-bath coupling strength,
bath memory, temperature, and spatial correlations and dissipation;
the light-harvesting antenna factor: electronic state
initialization; and reaction center factors: exciton trapping rate
and location. Next, we investigated the performance and sensitivity
of the FMO with small and large variations in its internal
parameters: chromophoric spatial locations, dipole moment
orientations, site energies, and chromophoric density. The exciton
trapping process in FMO complex shows to be significantly robust
with respect to perturbations in dipole moment orientations and site
energies due to its compact structure.

We believe that a key design principle for achieving optimal
and robust quantum transport performance is to allow for the
convergence of energy/time scales for the contributing internal and
external parameters. This phenomenon can be regarded as an examples
of a Goldilocks principle in the quantum regime.

\begin{acknowledgments}
We thank A. Aspuru-Guzik, A. Ishizaki, M. Sarovar, K. B. Whaley, for
useful discussions. We also thank J. H. Choi and D. Hayes for
helping us with extracting the FMO data. We acknowledge funding from
DARPA under the QuBE program (MM, AS, SL, HR), ENI (MM,SL), NSERC
(MM) and NSF (SL, AS,HR), and ISI, NEC, Lockheed Martin, Intel (SL).

\end{acknowledgments}

\appendix

\begin{figure}[tp]
\includegraphics[width=9cm,height=6cm]{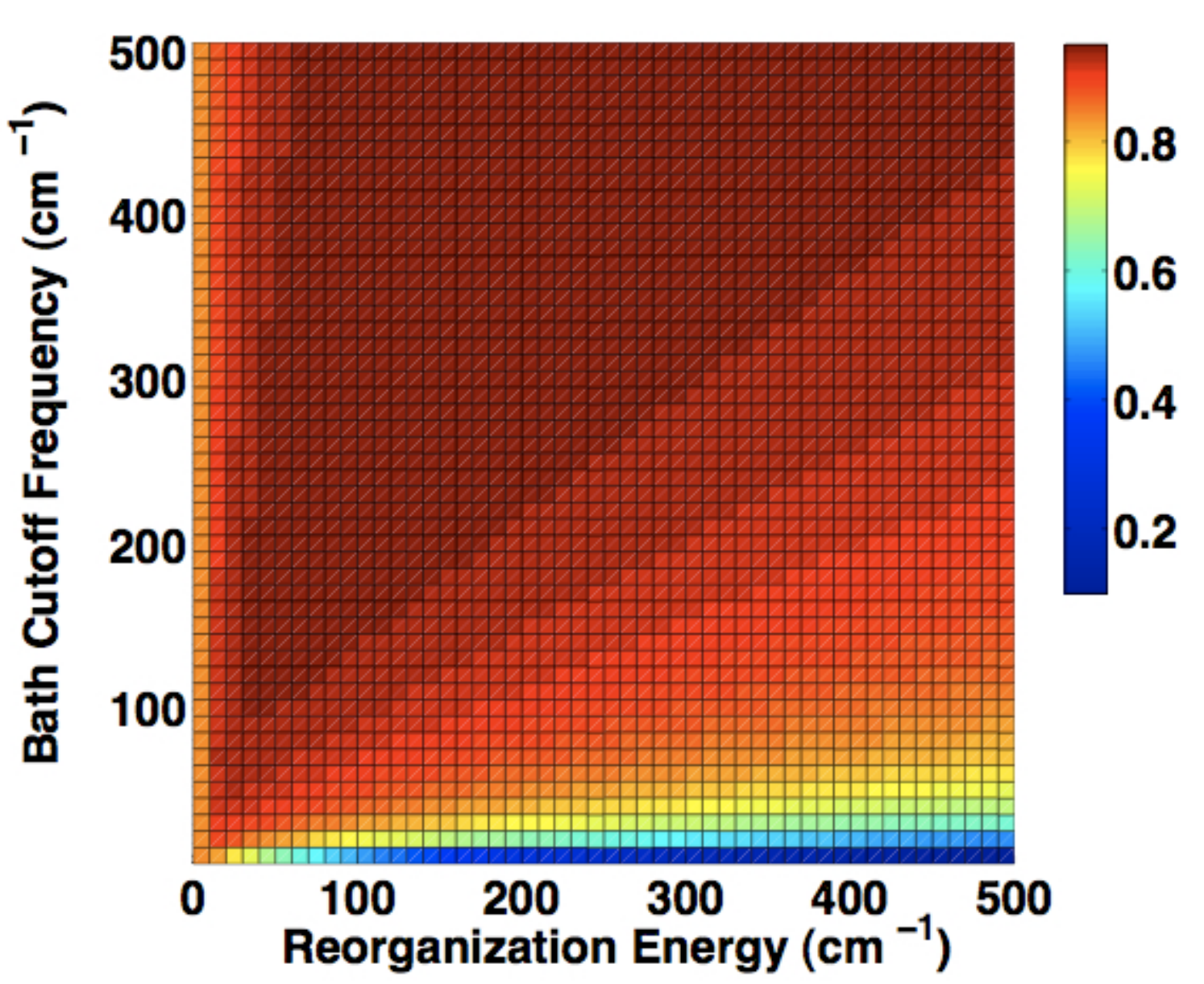}
\caption{The top view of FMO energy transfer landscape in the
presence of an Ohmic model of bath spectral density as a function of
reorganization energy and bath frequency cutoff at room temperature.
The ETE predicted by this model is below the estimate values by a
Lorentzian model and therefore it cannot explain near ideal
performance of FMO complex in exciton transport. Nevertheless, here
we can also observe the three distinct regimes of weak and strong
localizations close to $\gamma$ and $\lambda$ axes, and the
intermediate optimal ENAQT, separated by straight lines. This plot
confirms the role of parameter $\lambda T/\gamma$ as an effective
governing parameter.} \label{plotohm}
\end{figure}

\section{FMO free Hamiltonian}

In this work we use the following free Hamiltonian for the FMO
complex, given in Ref.~\cite{Cho05}:
\[
H =
 {\begin{pmatrix}
 280 & -106 & 8 & -5 & 6 & -8 & -4  \\
 -106 & 420 & 28 & 6 & 2 & 13 & 1  \\
 8 & 28 & 0 & -62 & -1 & -9 & 17\\
 -5 & 6 & -62 & 175 & -70 & -19 & -57\\
 6 & 2 & -1 & -70 & 320 & 40 & -2\\
 -8 & 13 & -9 & -19 & 40 & 360 & 32\\
 -4 & 1 & 17 & -57 & -2 & 32 & 260
 \end{pmatrix} }
\]

The estimated values of dipole moment orientations and positions of
the Mg atoms, representing the location of Bchls, are extracted from
the pdb file of the FMO complex \cite{Chopriv}. These data can be summarized in the following table:

\begin{figure}[tp]
\includegraphics[width=9cm,height=6cm]{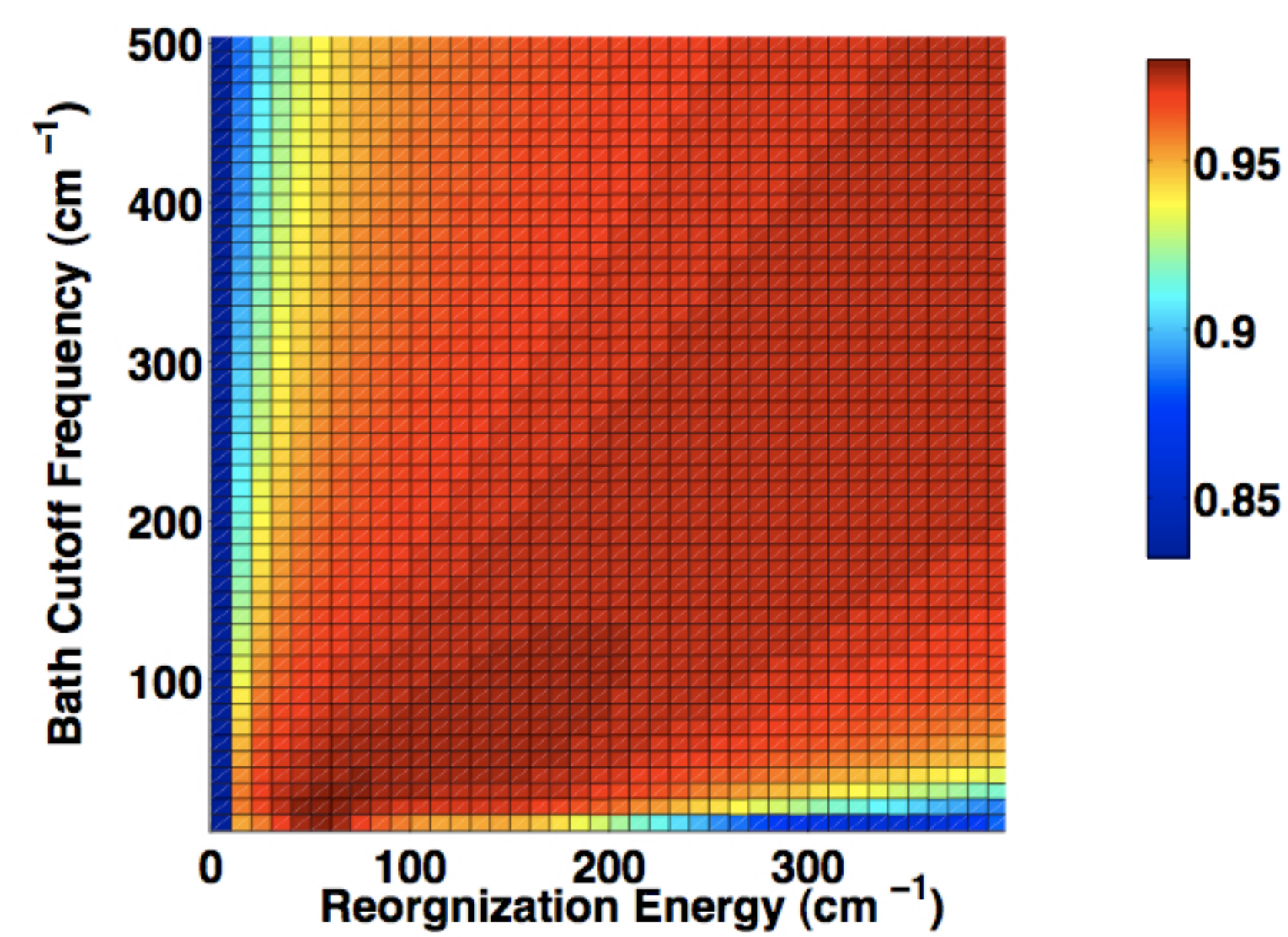}
\caption{Top view of the ETE landscape at $77^\circ K$ for a bath
with a Lorentzian spectral density. The general pattern of an
optimal transport region in between two limits of localized
low-efficient transport regimes, divided by rather straight lines,
is observed here as well. However, due to low-temperature effects
captured in the parameter $\lambda T/\gamma$, the adversarial high
reorganization and slow bath effects become less pronounced leading
to a higher ETE than those in Fig. \ref{topview} in the
corresponding limits.} \label{topview77}
\end{figure}

\begin{table}[ht]
\caption{Spatial location of Bchls and their dipole moment
orientation.} \centering
\begin{tabular}{c c c c c c}
\hline\hline Bchl & x ($\AA$) & y ($\AA$) & z ($\AA$) & $\theta$ &
$\phi$ \\ [0.5ex] \hline
 1 & 28.032 & 163.534 & 94.400 &  0.3816 & -0.6423+$\pi$ \\ \hline
    2 & 17.140 & 168.057& 100.162& 0.067& 0.5209+$\pi$ \\ \hline
     3 &5.409& 180.553&  97.621& 0.1399& 1.3616+$\pi$ \\ \hline
    4&  9.062& 187.635 & 89.474 &  0.257 &-0.6098+$\pi$\\ \hline
    5&  21.823& 185.260 & 84.721& -0.1606& 0.6899+$\pi$  \\ \hline
    6&  23.815 &173.888&  82.810&  -0.4214& -1.4686+$\pi$ \\ \hline
    7&  12.735& 174.887&  89.044& 0.578& -1.0076+$\pi$ \\ [1ex]
\hline
\end{tabular}
\label{fmo_data}
\end{table}
The Bchl-Bchl coupling in FMO is dipole-dipole interaction
\begin{eqnarray}
J_{jk}=\frac{C}{R_{jk}^{3}}(\mathbf{\mu }_{j}\cdot \mathbf{\mu }%
_{k}-\frac{3}{R_{jk}^{2}}(\mathbf{\mu }_{j}\cdot \mathbf{R}_{jk})(\mathbf{%
\mu }_{k}\cdot \mathbf{R}_{jk})),
\end{eqnarray}
for which we choose the constant $C|\mu|^2=134000$ $cm^{-1}\AA^3$ \cite{Renger06}.

\section{Energy transfer efficiency for Ohmic bath and cryogenic temperature}

Here, we demonstrate that separation of ETE landscape, as a function
of the parameter $\lambda T/\gamma$, into the
various regions with distinct quantum transport efficiencies is not
a mere property of either Lorentzian bath or high temperature limit.
A top view of FMO energy transfer landscape is shown for an Ohmic
spectral density at $298^\circ K$ and Lorentzian spectral density
at $77^\circ K$ in Fig. \ref{plotohm} and Fig. \ref{topview77}
respectively. In each figure, one can distinguish three different
regions including low-efficient weakly localized limit, optimal
EANQT, and low-efficient strongly localized limit, similar to the
results presented in Fig. \ref{topview}.

It should be noted that a bath with an Ohmic regularized spectral
density, $J(\omega)=\lambda(\omega/\gamma)\exp(-\omega/\gamma)$, has
often been employed in modeling the effect of bath fluctuations on
the spectroscopic readout of an FMO sample \cite{Fleming96,Cho05}.
Figure \ref{plotohm} presents the top view ETE landscape as a
function of bath reorganization energy and cutoff frequency with an
Ohmic density. The observed high efficient region stays lower than
its Lorentzian counterpart in Fig. \ref{topview}, therefore is
unable to explain the high efficiency of the FMO complex. This can
be seen as a confirmation for the theoretical modeling of the
solvent-protein environment with Lorentzian spectral densities
\cite{McKenzie}.


\begin{thebibliography}{99}
\bibitem{FMOpara} In this work, whenever it is not specified otherwise,
the environmental parameters for the FMO complex are chosen
according to the estimated values of reorganization energy $35
cm^{-1}$, bath cutoff frequency $50$ $cm^{-1}$, temperature $298$
$^\circ K$, trapping rate of $1$ $ps$, exciton life-time of $1$
$ns$.
\end{thebibliography}
\end{document}